\documentclass[preprint,12pt,authoryear]{elsarticle}

\usepackage{amssymb}
\usepackage{amsmath}
\usepackage{mathtools}

\let\pdv\relax
\usepackage{physics}
\usepackage{hyperref, cleveref}
\usepackage{epstopdf, epsfig}
\usepackage{amsfonts, bm, xcolor, subcaption}

\journal{International Journal of Multiphase Flow}

\begin{document}

\hfuzz=1pt

\begin{frontmatter}

\title{Effect of viscoelasticity on electrohydrodynamic drop deformation}

\author[label1]{Santanu Kumar Das}
\author[label2]{Sarika Shivaji Bangar}
\author[label1]{Amaresh Dalal}
\author[label2]{Gaurav Tomar}
\ead{gtom@iisc.ac.in}

\affiliation[label1]{organization={Department of Mechanical Engineering},
            addressline={Indian Institute of Technology Guwahati}, 
            city={Guwahati},
            postcode={781039}, 
            state={Assam},
            country={India}}

\affiliation[label2]{organization={Department of Mechanical Engineering},
            addressline={Indian Institute of Science}, 
            city={Bangalore},
            postcode={560012}, 
            state={Karnataka},
            country={India}}

\begin{abstract}
The impact of viscoelasticity on drop deformation in the presence of an electric field is investigated using both analytical and numerical methods. The study focuses on two configurations: a viscoelastic drop suspended in a Newtonian fluid and a Newtonian drop suspended in a viscoelastic medium. Oldroyd-B constitutive equation is employed to model constant viscosity viscoelasticity. Effect of Deborah number (ratio of polymer relaxation time to convective time scale) on drop deformation is studied and explained by examining the electric, elastic and viscous stresses at the interface. For small deformations, we apply the method of domain perturbations, and show that the viscoelastic properties of the drop significantly influence its deformation more than when the surrounding fluid is viscoelastic. Numerical computations are performed using a finite volume framework for larger drop deformations. The transient dynamics of the drops show distinct oscillatory patterns before eventually stabilizing at a steady deformation value. We observe a trend of decreased deformation in both configurations as the Deborah number increases. Relative magnitude of normal and tangential stresses plays a crucial role in drop deformation.
\end{abstract}

\begin{graphicalabstract}
\includegraphics[width=0.95\textwidth]{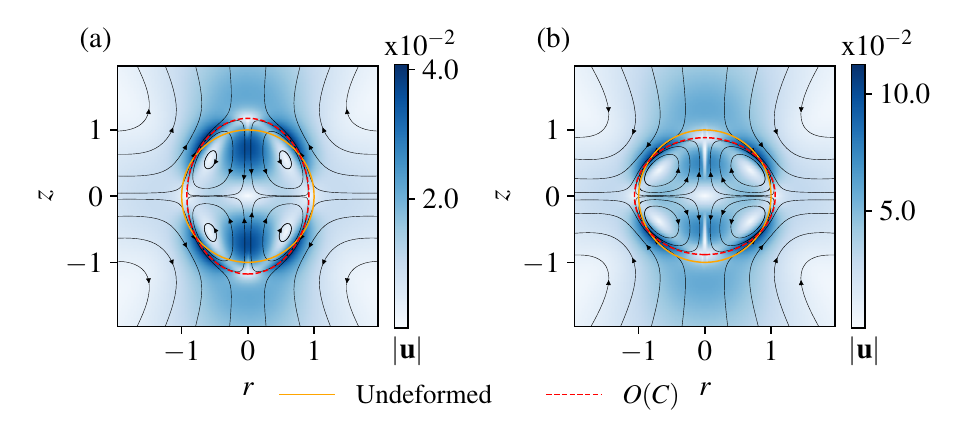}
\end{graphicalabstract}

\begin{highlights}
\item An analytical expression for the electrohydrodynamic deformation accounting for viscoelasticity.
\item Detailed numerical simulations of electrohydrodynamic deformation using volume of fluid method.
\item Polymeric stresses affect a viscoelastic drop in a Newtonian ambient more than a Newtonian drop in a viscoelastic ambient. 
\end{highlights}


\end{frontmatter}


\section{Introduction}

Viscoelastic fluids exhibit both viscous and elastic properties. These fluids usually consist of long chains of high molecular weight polymer molecules and can adopt various configurations and orientations. The interaction between these polymer chains and the surrounding fluid creates a complex internal micro-structure in the polymeric liquid, which gives rise to its characteristic flow behavior. Understanding polymer blends is crucial in industries such as food, paint, cosmetics, and processing, as the properties of these emulsions are directly linked to the micro-structure of the polymers resulting from drop deformation, coalescence, and breakup. In dilute emulsions, the dynamics of a single drop provides complete information about its behavior. As a result, extensive studies have been carried out to investigate drop deformation and breakup in simple linear flows \citep{Taylor1932}. Most of these emulsion studies were restricted to both phases being Newtonian \citep{Stone1994}. However, non-Newtonian fluids can significantly modify the dynamics of drop deformation and breakup \citep{Tucker2002,Yue2005}.

In most studies on non-Newtonian emulsions, the focus has been on the effect of shear and extensional flows. The dynamics of droplets in Newtonian systems is governed by the viscosity ratio and the capillary number, which represents the ratio of the viscous stretching force to the interfacial resistive force. There exists a critical capillary number (ratio of shear to capillary forces) below which droplets maintain their elongated shapes and above which they break. When viscoelasticity is introduced, it alters the extent of drop deformation and the critical capillary number for breakup. This effect is attributed to non-zero normal stresses, changes in viscous and viscoelastic stresses, and flow modification \citep{Ramaswamy1999,Ramaswamy1999a,Hooper2001}.

Compared to a Newtonian drop, the deformation of a viscoelastic drop decreases due to the presence of normal stresses in the drop phase. \citet{Gauthier1971} and \citet{Varanasi1994} found that a higher critical capillary number is required for breakup of viscoelastic drops in a Newtonian matrix in comparison to that for Newtonian drops in a Newtonian matrix. \citet{Flumerfelt1972} performed experiments with a Newtonian drop sheared in a viscoelastic matrix and showed that the critical shear rate increases with increasing matrix elasticity. \citet{Toose1995} performed simulations of viscoelastic drop deformation, using a boundary integral method, and showed that for small deformations, drop viscoelasticity affects only the transient behavior. In systems where either the drop or the matrix phase is viscoelastic, they showed that drop elasticity inhibits deformation. In contrast, matrix elasticity enhances drop deformation in a simple shear flow \citep{Elmendorp1986}.  Similar conclusions were drawn from experiments with Boger fluids \citep{Mighri1997,Mighri1998}. However, there are contradictory findings, with some studies indicating that matrix viscoelasticity impedes drop breakup in simple shear flows \citep{Guido2003,Sibillo2004}. This contradictory behavior was examined by \citet{Yue2005}, who reported that the deformation of a Newtonian drop in a viscoelastic matrix is non-monotonic. This study was conducted using a diffuse interface method. Although there have been several studies on the effect of shear and extensional flows of viscoelastic emulsions, the effect of viscoelasticity in electrohydrodynamic flows has been scantily studied.

Electric field can be employed for controlling emulsions to achieve desired properties. Electric field can cause droplets to deform and breakup by inducing shear stress at the droplet interface. \citet{OKonski1957} investigated the extent of droplet deformation in a uniform electric field by defining free energy for the system using electrostatic and interfacial energy and assuming the droplet to be ellipsoidal. \citet{Taylor1966} calculated the flow in the drop and in the suspending fluid due to the shear forces induced by the electric field forces at the drop interface. Corrections to the drop shape in second order in electric capillary number were computed by \citet{Ajayi1978}. \citet{Torza1971} studied the deformation of the drop under the influence of an alternating electric field up to first order in electric capillary number. All these studies analyze Newtonian drops in a Newtonian ambient, but there is limited analysis for viscoelastic drops subjected to a uniform electric field. \cite{Ha1999} conducted an asymptotic analysis in the limit of a small Deborah number calculating the change in the drop shape. \citet{Ha2000} experimentally observed the effects of viscoelasticity on the deformation and breakup of a drop in a uniform electric field, concluding that the critical electric field strength for drop breakup increases with viscoelasticity. \citet{Lima2014} performed numerical simulations using a finite volume method and found that the drop deformation decreases with an increase in the relaxation time of the viscoelastic fluid modeled using Giesekus constitutive model. They found that the electric field magnitude primarily influences the deformation but other parameters such as the ratio of electrical conductivity and permittivity also play an important role.

In this work, we study drop deformation of an Oldroyd-B fluid at a finite Reynolds number. We consider the drop or the matrix to be viscoelastic. The Oldroyd-B model is chosen for its shear-independent viscosity and ability to sustain a positive first normal stress and a zero second normal stress. It is a quasi-linear constitutive model with a single relaxation time. We consider both phases to be leaky-dielectric. In particular, we focus on the role of viscoelastic stresses in two different configurations, namely, viscoelastic drops in a Newtonian matrix and Newtonian drops in a viscoelastic matrix.

In what follows, we first discuss the governing equations and boundary conditions in section \ref{sec:formulation}, asymptotic analysis in sec.\ref{sec:analytical}. Numerical formulation and results are given in secs.\ref{sec:numerical}. Finally, important conclusions from the study are presented in sec.\ref{sec:conclusions}.

\section{Problem Formulation} \label{sec:formulation}

We study the deformation of a neutrally buoyant drop of radius $a$ suspended in a medium in the presence of a uniform electric field $\boldsymbol{E}_\infty$. Both the drop and the ambient phases are modeled as leaky dielectric and the viscoelastic behavior is captured using the Oldroyd-B model (polymeric solution). The viscosity, electric permittivity, and electric conductivity are denoted using $\mu$, $\varepsilon$, and $\sigma$, respectively. $\mu_s$ and $\mu_p$ are the solvent and polymer viscosity, respectively, of the viscoelastic fluid with total viscosity as $\mu = \mu_s + \mu_p$. The fluids involved are assumed to be incompressible with a constant surface tension coefficient, $\gamma$, at the interface between the two fluids. The subscripts $i$ and $e$ identify the quantities internal and external to the drop, respectively.

\begin{figure}[h]
    \centerline{\includegraphics[width=0.5\textwidth]{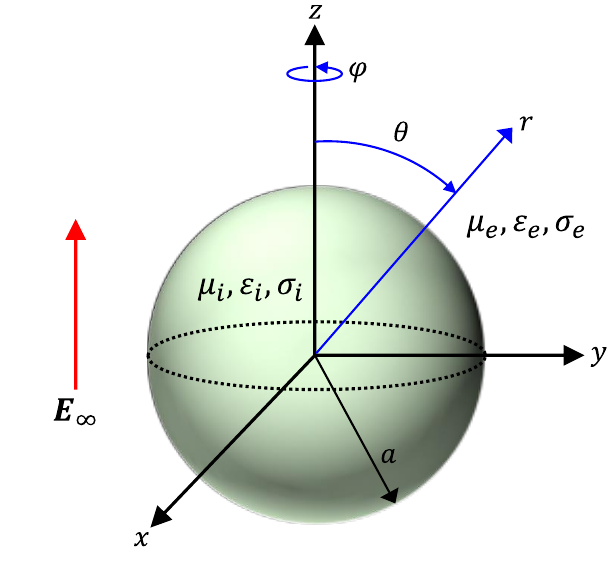}}
    \caption{Schematic representation of a drop of radius $a$ in the presence of an imposed uniform electric field $\boldsymbol{E}_\infty$. A spherical coordinate system ($r, \theta, \varphi$) is considered which is attached to the centroid of the drop.}
    \label{fig:1_schematic_theoretical}
\end{figure}

We use the following scaling parameters for nondimensionalization: length $\sim a$ (radius of the drop), electric field $\sim E_\infty$, pressure and the polymeric stress $\sim \mu_e U/a$ (where $U$ is the characteristic velocity scale), electric potential $\sim a E_\infty$, and electric stress $\sim \varepsilon_e E_\infty^2$. Since there is no inherent velocity scale, characteristic velocity can be determined by the balance between the electric and viscous stresses as $U \sim a \varepsilon_e E_\infty^2 / \mu_e$. The following fluid property ratios and dimensionless parameters are defined using the above scaling: density ratio $\rho_r = \rho_i/\rho_e$, viscosity ratio $\mu_r = \mu_i / \mu_e$, conductivity ratio $R = \sigma_i / \sigma_e$, permittivity ratio $Q = \varepsilon_i / \varepsilon_e$, Reynolds number $Re = \rho a U / \mu_e$, electric capillary number $C = \varepsilon_e a E_\infty^2 / \gamma$, and electric Reynolds number $Re_E = \varepsilon_e U / a \sigma_e$. $Re$ represents the ratio of the inertia forces to the viscous forces, and $C$ represents the ratio of the electrostatic forces to the surface tension forces. The term $Re_E$ signifies the ratio of the charge relaxation time scale ($\varepsilon_e/\sigma_e$) to the flow time scale ($U/a$). A low $Re_E$ indicates fast relaxation of the charge with respect to the charge advection. Thus, for low $Re_E$ a separate charge conservation equation need not be solved. Furthermore, two additional dimensionless parameters are involved in studying viscoelastic fluids. The ratio of the solvent viscosity to the total viscosity of the phase is denoted by $\beta = \mu_s / (\mu_s + \mu_p) = \mu_s / \mu$. Deborah number $De( = \lambda U/a)$ represents the ratio of the relaxation time scale $(\lambda)$ to the flow time scale $(a/U)$. Deborah number indicates how a particular material will behave over a given time frame. If the observation time is long or the relaxation time of the material is short, then the material is expected to behave like a fluid. On the other hand, if the relaxation time of the material is large or the observation time is short, the material will have a high Deborah number and behave like an elastic solid.

\subsection{Governing equations and boundary conditions} \label{sec:equations}

Under the assumption of electrostatics, we can represent the electric field as a gradient of electric potential ($\phi$). The Laplace equations govern the electric potential both inside and outside the drop, given by
\begin{equation}
    \grad^2 {\phi_i} = 0 \; , \quad \grad^2 {\phi_e} = 0 \; ,
\end{equation}
where $\phi_i$ and $\phi_e$ represent the electric potential inside and outside the drop, respectively. The electric potential datum is set to zero ($\phi_i = 0$) at $r = 0$. Symmetry boundary condition is imposed at the center of the drop as,
\begin{equation}
    \pdv{\phi_i}{r} = 0 \quad \text{at} \quad r = 0 \; .
\end{equation}
The electric field outside the drop should approach the imposed electric field far away from the drop, which can be expressed as:
\begin{equation}
    \phi_e = \boldsymbol{E}_\infty \cdot x = r \cos \theta \quad \text{as} \quad r \rightarrow \infty \; .
\end{equation}
The electric potential is continuous at the drop interface:
\begin{equation}
    \phi_i = \phi_e \quad \text{at} \quad r = 1 + f(\theta) \; ,
\end{equation}
where $f(\theta)$ is a function that describes the shape of the drop interface. Additionally, at the drop interface, the conduction current is continuous, and it is given by
\begin{equation}
    R~\boldsymbol{E}_i \cdot \vb{n} = \boldsymbol{E}_e \cdot \vb{n} \quad \text{at} \quad r = 1 + f(\theta) \; ,
\end{equation}
where $\vb{n}$ is the unit normal at the drop interface pointing outwards. The stress due to the electric field can be calculated using the Maxwell stress tensor $(\boldsymbol{\tau}^E)$, ignoring the effects of the magnetic field. The electric stress inside and outside of the drop is given as
\begin{align}
\begin{rcases}
    \boldsymbol{\tau}_i^E = Q &\left( \boldsymbol{E}_i \boldsymbol{E}_i - \frac{1}{2} \boldsymbol{E}_i \cdot \boldsymbol{E}_i \boldsymbol{I} \right) \; , \\
    \boldsymbol{\tau}_e^E = &\left( \boldsymbol{E}_e \boldsymbol{E}_e - \frac{1}{2} \boldsymbol{E}_e \cdot \boldsymbol{E}_e \boldsymbol{I} \right) \; .
\end{rcases}
\end{align}
The electric stress can be divided into normal and tangential components. The tangential component is responsible for fluid flow, while the imbalance of the normal stress causes the drop to deform. Additionally, the hydrodynamic stress generated by the fluid flow is discontinuous at the interface. To accurately determine the deformation of the drop, it is necessary to solve the fluid flow equations.

The mass conservation equation for incompressible flow is as follows
\begin{align}
\begin{rcases}
    \div{\vb{u}_i} = 0 \; , \\
    \div{\vb{u}_e} = 0 \; ,
\end{rcases}
\end{align}
while the momentum conservation is given by the Navier-Stokes equation as
\begin{align}
\begin{rcases}
    Re\left(\pdv{\vb{u}_i}{t} + \vb{u}_i.\grad\vb{u}_i\right) &= -\grad p_i + \grad.\bm{\tau}_i  \; , \\
    Re\left(\pdv{\vb{u}_e}{t} + \vb{u}_e.\grad\vb{u}_e\right) &= -\grad p_e + \grad.\bm{\tau}_e  \; ,
\end{rcases}
\end{align}
where $p$ and $\vb{u}$ represent the pressure field and velocity field, respectively. Here, the characteristic viscous scale $(\mu_e U/a)$ has been used for the non-dimensionalization of pressure and viscous stress since flow is dominated by viscous effects.

While accounting for viscoelasticity, the hydrodynamic stress is split into solvent and polymeric components as $\boldsymbol{\tau}^H = \boldsymbol{\tau}^{sH} + \boldsymbol{\tau}^{pH}$. The Newtonian constitutive relation governs the solvent component of the hydrodynamic stress as
\begin{align}
\begin{rcases}
    \boldsymbol{\tau}^{sH}_i = \mu_r \beta_i &\left( \grad{\vb{u}_i} + \left( \grad{\vb{u}_i} \right)^T \right) \; , \\
    \boldsymbol{\tau}^{sH}_e = \beta_e &\left( \grad{\vb{u}_e} + \left( \grad{\vb{u}_e} \right)^T \right) \; ,
\end{rcases}
\end{align}
while the polymeric stress is governed by the Oldroyd-B constitutive model given as
\begin{align}
\begin{rcases}
    \boldsymbol{\tau}^{pH}_i + \Lambda_i De \bigg( \frac{\partial \boldsymbol{\tau}^{pH}_i}{\partial t} + \vb{u}_i \cdot \grad{\boldsymbol{\tau}^{pH}_i} - (\grad{\vb{u}_i}) \boldsymbol{\tau}^{pH}_i &- \boldsymbol{\tau}^{pH}_i (\grad{\vb{u}_i})^T \bigg) \\
    &= \mu_r (1 - \beta_i) \left( \grad{\vb{u}_i} + \left( \grad{\vb{u}_i} \right)^T \right) \; , \\
    \boldsymbol{\tau}^{pH}_e + \Lambda_e De \bigg( \frac{\partial \boldsymbol{\tau}^{pH}_e}{\partial t} + \vb{u}_e \cdot \grad{\boldsymbol{\tau}^{pH}_e} - (\grad{\vb{u}_e}) \boldsymbol{\tau}^{pH}_e &- \boldsymbol{\tau}^{pH}_e (\grad{\vb{u}_e})^T \bigg) \\
    &= (1 - \beta_e) \left( \grad{\vb{u}_e} + \left( \grad{\vb{u}_e} \right)^T \right) \; ,
\end{rcases}
\end{align}
where $\beta_i(= \mu_{is} / \mu_i)$ and $\beta_e(= \mu_{es} / \mu_e)$ are the ratio of the solvent viscosity to the total viscosity of the respective phases. Additionally, $\Lambda_i(= \lambda_i / \lambda)$ and $\Lambda_e(= \lambda_e / \lambda)$ with $\lambda = \lambda_i + \lambda_e$ are the respective time relaxation constant of both the phases.

The fluid flow must comply with the following boundary conditions. The velocity field inside the drop ($\vb{u}_i$) is bounded at the center of the drop ($r=0$). The symmetry of the velocity field at the drop center is given by,
\begin{equation}
    \pdv{\vb{u}_i}{r} = 0 \quad \text{at} \quad r = 0 \; .
\end{equation}
The velocity field outside the drop diminishes far away from the drop
\begin{equation}
    \bm{u}_e \to 0 \quad \text{as} \quad r \rightarrow \infty \; .
\end{equation}
At the drop interface, the velocity field is continuous, and the normal component of the velocity field vanishes.
\begin{align}
\begin{rcases}
    \vb{u}_i \cdot \vb{n} = \vb{u}_e \cdot \vb{n} \quad &\text{at} \quad r = 1 + f(\theta) \; , \\
    \vb{u}_i = \vb{u}_e \quad &\text{at} \quad r = 1 + f(\theta) \; .
\end{rcases}
\end{align}
The tangential component of the total stress (hydrodynamic and electrical Maxwell stresses) is continuous at the drop surface given as
\begin{equation}
    \vb{n} \cdot (\boldsymbol{\tau}_i^H + \boldsymbol{\tau}_i^E) \cdot (\vb{I} - \vb{n}\vb{n}) = \vb{n} \cdot (\boldsymbol{\tau}_e^H + \boldsymbol{\tau}_e^E) \cdot (\vb{I} - \vb{n}\vb{n}) \quad \text{at} \quad r = 1 + f(\theta) \; .
\end{equation}
Furthermore, the normal component of the total stress is balanced by the capillary stress induced by surface tension as,
\begin{equation}
    \vb{n} \cdot (\boldsymbol{\tau}_e^H + \boldsymbol{\tau}_e^E) \cdot \vb{n} - \vb{n} \cdot (\boldsymbol{\tau}_i^H + \boldsymbol{\tau}_i^E) \cdot \vb{n} = \frac{1}{C} (\div{\vb{n}}) \quad \text{at} \quad r = 1 + f(\theta) \; .
\end{equation}

\section{Asymptotic analysis for small shape deformation} \label{sec:analytical} 

The problem formulated in the previous section is nonlinear despite the governing equations being linear, primarily due to the nonlinear boundary conditions. Another difficulty inherent in this problem is that the location of the interface where the boundary conditions apply is initially unknown and needs to be determined as part of the solution. Therefore, numerical methods are necessary for solving significant deformations at arbitrary electric capillary numbers. However, for small deformations, an analytical solution can be obtained through asymptotic expansion for small values of electric capillary number. We present the analysis of a viscoelastic drop in a viscoelastic ambient subjected to an external electric field. For this, we use a spherical coordinate system ($r, \theta, \varphi$) with the origin at the center of the drop. The schematic of the system is shown in \Cref{fig:1_schematic_theoretical}. The angle $\theta$ is measured from the positive $z$ axis, assumed to be the axis of symmetry. The applied electric field is in the direction of the axis of symmetry ($\boldsymbol{E}_\infty = E_\infty \boldsymbol{e}_z$). We make the Stokes flow assumption to simplify the flow governing equations to the following form
\begin{align}
\begin{rcases}
    \grad{p_i} = \grad.\boldsymbol{\tau}_i & \; , \quad \div{\vb{u}_i} = 0 \; , \\
    \grad{p_e} = \grad.\boldsymbol{\tau}_e & \; , \quad \div{\vb{u}_e} = 0 \; .
\end{rcases}
\end{align}

\subsection{Expansion in terms of \texorpdfstring{$C$}{C} and \texorpdfstring{$De$}{De}}

We employ the domain perturbation procedure by considering the asymptotic limit when $C\ll1$. This expansion allows for a small deviation of the interface shape from the spherical shape and linearization of the boundary conditions at the deformed interface to those at the spherical interface. Additionally, the effect of viscoelasticity can be determined through asymptotic expansion for small values of $De$. Therefore, a two-parameter asymptotic expansion is employed considering $C \ll 1$, and $C^2,CDe,De^2 \ll De$. The physical quantities are expanded using $C$ and $De$ as perturbation parameters in the following form
\begin{align}
\begin{rcases}
    \vb{u}_{i,e} &= \vb{u}_{i,e}^{(0)} + C~\vb{u}_{i,e}^{(C)} + De~\vb{u}_{i,e}^{(De)} + C De~\vb{u}_{i,e}^{(CDe)} + \cdots \; , \\
    p_{i,e} &= p_{i,e}^{(0)} + C~p_{i,e}^{(C)} + De~p_{i,e}^{(De)} + C De~p_{i,e}^{(C De)} + \cdots \; , \\
    \boldsymbol{E}_{i,e} &= \boldsymbol{E}_{i,e}^{(0)} + C~\boldsymbol{E}_{i,e}^{(C)} + De~\boldsymbol{E}_{i,e}^{(De)} + C De~\boldsymbol{E}_{i,e}^{(C De)} + \cdots \; , \\
    \boldsymbol{\tau}_{i,e} &= \boldsymbol{\tau}_{i,e}^{(0)} + C~\boldsymbol{\tau}_{i,e}^{(C)} + De~\boldsymbol{\tau}_{i,e}^{(De)} + C De~\boldsymbol{\tau}_{i,e}^{(C De)} + \cdots \; ,
\end{rcases}
\end{align}
where $\boldsymbol{\tau}_{i,e} = \boldsymbol{\tau}_{i,e}^H + \boldsymbol{\tau}_{i,e}^E$. The drop remains spherical when $C$ is zero. For small values of $C$, the Newtonian velocity field causes the drop to deviate from the spherical shape. Therefore, the interface shape is given by the following expression
\begin{equation}
\begin{split}
    r_s = 1 &+ C \{f^{(C)}(\theta) + Def^{(C De)}(\theta) + O(De^2)\} \\
    &+ C^2\{f^{(C^2)}(\theta) + Def^{(C^2 De)}(\theta) + O(De^2)\} + O(C^3) \; .
\end{split}
\end{equation}

In the presence of an electric field, the spherical shape of the interface of the drop becomes deformed. However, since the exact shape of the deformed interface is unknown, it is necessary to implement interface boundary conditions at the unknown interface and solve for the interface as part of the solution. If the surface is represented by the equation $F = r_s - (1 + f(\theta))$, the normal to the surface can be calculated as $\mathbf{n} = \nabla F / \lvert \nabla F \rvert$. The divergence of the normal vector gives the curvature of the surface as
\begin{equation}
\begin{split}
    \kappa = \grad \cdot \mathbf{n} &= 2 - C \Big[~2 f^{(C)} + \cot\theta~f^{\prime (C)} + f^{\prime \prime (C)} + De~ \big(~2f^{(C De)} + \cot\theta~f^{\prime (C De)}\\
    &+ f^{\prime \prime (C De)} \big) + O(De^2) \Big] + C^2 \Big[~2 (f^{(C)})^2 - 2 f^{(C^2)} + 2 \cot\theta~f^{(C)}f^{\prime (C)}\\
    &- 2 \cot\theta~f^{\prime (C^2)} + 2 f^{(C)}f^{\prime \prime (C)} - f^{\prime \prime (C^2)} + O(De) \Big] + O(C^3) \; .
\end{split}
\end{equation}

\subsection{Method of domain perturbations}

The boundary conditions at the interface involve both normal and tangential components of velocity and stress. All vectors and tensors are represented in a spherical axisymmetric coordinate system. The components of vectors and tensors, normal or tangential to the deformed surface, are calculated by rotating them with a rotation matrix through an angle denoted by $\alpha$, which is the angle between the normal to the deformed surface and the normal to the spherical surface. Since $\alpha$ is typically small, we can approximate $\sin\alpha \approx \alpha$ and $\cos\alpha \approx 1$. If $\boldsymbol{a}$ is any vector with $a_r$ and $a_\theta$ components in a spherical coordinate system, its components in the direction normal and tangential to the deformed surface are given by
\begin{align}
\begin{rcases}
    a_n &= a_r\cos\alpha - a_\theta \sin\alpha \approx a_r - a_\theta f^{\prime}(\theta) \; ,\\
    a_t &= a_r\sin\alpha + a_\theta\cos\alpha \approx a_\theta + a_r f^{\prime}(\theta) \; .
\end{rcases}
\end{align}
Using the same rotation matrix for the transformation of the stress tensor and using the mentioned approximations, components of the stress tensor are given by
\begin{align}
\begin{rcases}
    \tau_{nn} &= \tau_{rr} - 2 \tau_{r\theta} f^{\prime}(\theta) \; , \\
    \tau_{nt} &= \tau_{r\theta} + (\tau_{rr} - \tau_{\theta\theta}) f^{\prime}(\theta) \; , \\
    \tau_{tt} &= \tau_{\theta\theta} + 2 \tau_{r\theta} f(\theta) \; .
\end{rcases}
\end{align}

In order to implement boundary conditions at the interface, it is necessary to evaluate vector and tensor components at the interface, which are unknown. Therefore, these components are simplified to quantities that can be evaluated at the spherical surface using Taylor's series. By substituting these expanded physical quantities in the governing equations and boundary conditions, a system of equations and boundary conditions at orders $O(1)$, $O(C)$, and $O(De)$ are obtained, which leads to the correction of the drop shape to orders $O(C)$, $O(C^2)$ and $O(C De)$, respectively. The solution to the equations of various orders are given in \cref{appA}.

\subsection{Drop deformation}

When a drop and an ambient medium are conducting, a non-zero tangential stress is generated at the interface because of the electric field. This stress is balanced by viscous tangential stress, resulting in a circulatory flow pattern in both phases. The drop interface undergoes deformation at $O(C)$ to satisfy the normal stress condition at the interface. The electric field changes due to the deformation of the drop, which leads to corrections to the electric field and flow field at $O(C)$, causing the drop to deform to $O(C^2)$. The flow field at $O(1)$ gives rise to polymeric stress at $O(De)$, leading to corrections in the flow field at $O(De)$ and resulting in the deformation of the drop at $O(C De)$. The deformed interface of the drop is given by
\begin{equation}
    r_s(\theta) = 1 + C f^{(C)}(\theta) + C^2 f^{(C^2)}(\theta) + C De f^{(C De)}(\theta) \; ,
\end{equation}
where,
\begin{align}
\begin{rcases}
    f^{(C)}(\theta) &= B^{(C)}_S P_2(\cos\theta) \; , \\
    f^{(C^2)}(\theta) &= A^{(C^2)}_S + B^{(C^2)}_S P_2(\cos\theta) + C^{(C^2)}_S P_4(\cos\theta) \; , \\
    f^{(CDe)}(\theta) &= B^{(CDe)}_S P_2(\cos\theta) + C^{(CDe)}_S P_4(\cos\theta) \; .
\end{rcases}
\end{align}
At $O(C)$, the correction to the drop shape takes the form of the second mode of the Legendre polynomial. Meanwhile, at $O(C^2)$ and $O(C De)$, the deformation is represented by the second and fourth modes of the Legendre polynomial. Only even modes of the Legendre polynomial appear due to the symmetry of the problem. The expressions for the constants in the equations are detailed in \cref{appA}.

Now, we present the expressions for small deformation of the drop, which were obtained through asymptotic analysis in the limit of small $C$ and small $De$. The approximate equation for the deformation is
\begin{equation}
    D = C D^{(C)} + C^2 D^{(C^2)} + C De D^{(C De)} \; ,
\end{equation}
where
\begin{equation}
    D^{(C)} = \frac{9}{80} \frac{1}{(1 + \mu_r) (2 + R)^2} \Big[ (6 + 9 \mu_r) R -(16 + 19\mu_r)Q + 5 (1 + \mu_r) (1 + R^2) \Big] \; ,
\end{equation}
\begin{equation}
\begin{split}
    D^{(C^2)} &= \frac{D^{(C)}}{(2 + R)^3(1 + \mu_r)^2} \Bigg[\frac{(2743 + \mu_r (20616 + 15281 \mu_r))R}{8400}\\
    &+ \frac{(5120 + \mu_r (31413 + 24997 \mu_r)) R^2}{8400} + \frac{1}{80} (1 + \mu_r)^2 (-154 + 139 R^3)\\
    &+ Q \bigg\{ \frac{22096 + \mu_r (36627 + 15827 \mu_r)}{4200}\\
    &+ R \left(\frac{-50480 - \mu_r(122133 + 70357\mu_r)}{8400}\right) \bigg\} \Bigg] \; ,
\end{split}
\end{equation}
\begin{equation}
\begin{split}
    D^{(C De)} = \frac{27(Q - R)^2}{56000 (1 + \mu_r)^3 (2 + R)^4}
    \Big[ &\Lambda_e (1 - \beta_e)(408 - 443\mu_r)\\
    &- 4\Lambda_i \mu_r (1 - \beta_i)(787 + 998 \mu_r) \Big] \; .
    \label{eqn:D_CDe_full}
\end{split}
\end{equation}

\Cref{eqn:D_CDe_full} indicates that the contribution of viscoelasticity to the deformation is zero when $R=Q$. When $R<Q$, the magnitude of $D^{(CDe)}$ increases as $R$ decreases. To investigate the impact of viscoelasticity on drop deformation, we are considering two configurations: a viscoelastic drop in a Newtonian fluid ($V_dN_m$) and a Newtonian drop in a viscoelastic medium ($N_dV_m$). When the drop is viscoelastic and medium is Newtonian ($V_dN_m$), we have $\beta_e = 1$ and $\Lambda_i = 1$, and the correction term becomes
\begin{equation}
    D^{(C De)} = -\frac{27(Q - R)^2 (4 \mu_r (1 - \beta_i)(787 + 998 \mu_r))}{56000 (1 + \mu_r)^3 (2 + R)^4} \; .
\end{equation}
Since $(1 - \beta_i)$ is always positive, it can be seen that deformation $D^{(C De)}$ is negative, indicating that deformation decreases with Deborah number. For a viscosity ratio of 1, we obtain,
\begin{equation}
    D^{(C De)} = -\frac{1377(Q - R)^2 (1 - \beta_i)}{3200 (2 + R)^4}
    \label{DCaDe_M1_VN} \; .
\end{equation}

Switching the phases, when the ambient is viscoelastic, and the drop is Newtonian ($N_dV_m$), we have $\beta_i = 1$ and $\Lambda_e = 1$. In this case, the correction term takes the form
\begin{equation}
    D^{(C De)} = \frac{27(Q - R)^2(\Lambda_e(1 - \beta_e)(408 - 443\mu_r))}{56000 (1 + \mu_r)^3 (2 + R)^4} \; .
\end{equation}
Since $(1 - \beta_e)$ is always positive, it can be seen that deformation $D^{(C De)}$ can be positive or negative depending on the viscosity ratio. Keeping the viscosity ratio to 1, we obtain
\begin{equation}
    D^{(C De)} = -\frac{27(Q - R)^2(1 - \beta_e)}{3200 (2 + R)^4} \; .
    \label{DCaDe_M1_NV}
\end{equation}

The deformation of a drop always decreases with the Deborah number if either the drop or the ambient is viscoelastic. However, if drop deformation is oblate, the magnitude of deformation increases as oblate deformations are negative by definition. Thus, viscoelasticity increases the prolate deformation but decreases the magnitude of oblate deformation. When comparing \Cref{DCaDe_M1_VN,DCaDe_M1_NV}, it is evident that for a viscosity ratio of 1, the deviation of drop deformation is greater when the drop is viscoelastic compared to when the ambient is viscoelastic.

\begin{figure}
    \centering
    \includegraphics[width=\textwidth]{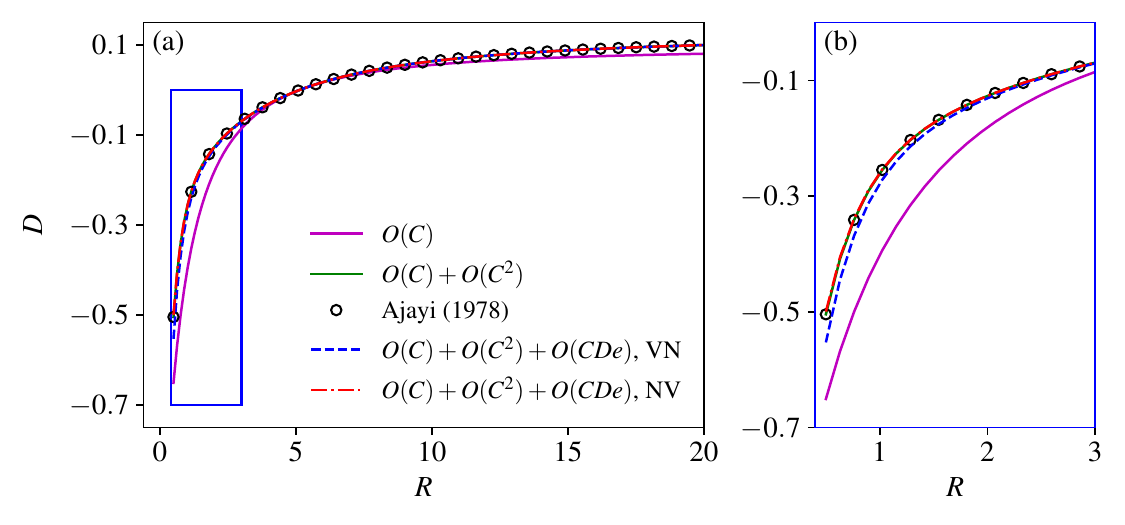}
    \caption{(a) Variation of drop deformation with $R$ for $Q=10$. The other dimensionless parameters considered are $C=0.2$, $De=0.2$, $\beta=0.1$, $Re=1$, and $\mu_r=1$. Markers show the $O(C) + O(C^2)$ deformation from \cite{Ajayi1978}. (b) Deviation from Newtonian behavior is higher for lower values of $R$.}
    \label{fig:2_deformation_vs_cond}
\end{figure}

\Cref{fig:2_deformation_vs_cond} illustrates the variation of drop deformation with the conductivity ratio ($R$) at a fixed permittivity ratio $Q=10$, viscosity ratio $\mu_r=1$ and electric capillary number $C=0.2$. The drop deformation is plotted for a Newtonian drop at $O(C)$ as analyzed by \citet{Taylor1966} and at $O(C^2)$ as calculated by \citet{Ajayi1978}. To investigate the effect of viscoelasticity, we consider two configurations: $V_dN_m$ and $N_dV_m$. In these cases, either the drop or the medium is viscoelastic, with a constant $De=0.2$. The ratio of solvent to total viscosity is set at $\beta=0.1$. \Cref{fig:2_deformation_vs_cond}(b) provides a magnified view of the region highlighted by the blue box in \Cref{fig:2_deformation_vs_cond}(a). This zoomed-in view specifically examines the behavior of the system for lower values of R while keeping Q fixed at 10. As predicted by the expression for $D^{(CDe)}$, the deviation from Newtonian behavior becomes more pronounced as the conductivity ratio decreases. Furthermore, we can see that the addition of the $O(C^2)$ correction significantly contributes to drop deformation. However, the contribution from the $O(CDe)$ term is relatively small. We can observe that the deformation for a Newtonian drop in a viscoelastic medium $(N_dV_m)$ closely overlaps with the Newtonian plot ($O(C) + O(C^2)$). In contrast, a slight deviation from Newtonian behavior is observed when the drop is viscoelastic but surrounded by a Newtonian ambient $(V_dN_m)$. This indicates that the viscoelasticity of the drop has a more significant impact on drop deformation compared to the viscoelasticity of the surrounding medium, as demonstrated in \Cref{DCaDe_M1_VN,DCaDe_M1_NV}.

\textbf{\begin{figure}[htbp]
  \centering
  \begin{minipage}{0.9\linewidth}
    \includegraphics[width=0.8\linewidth]{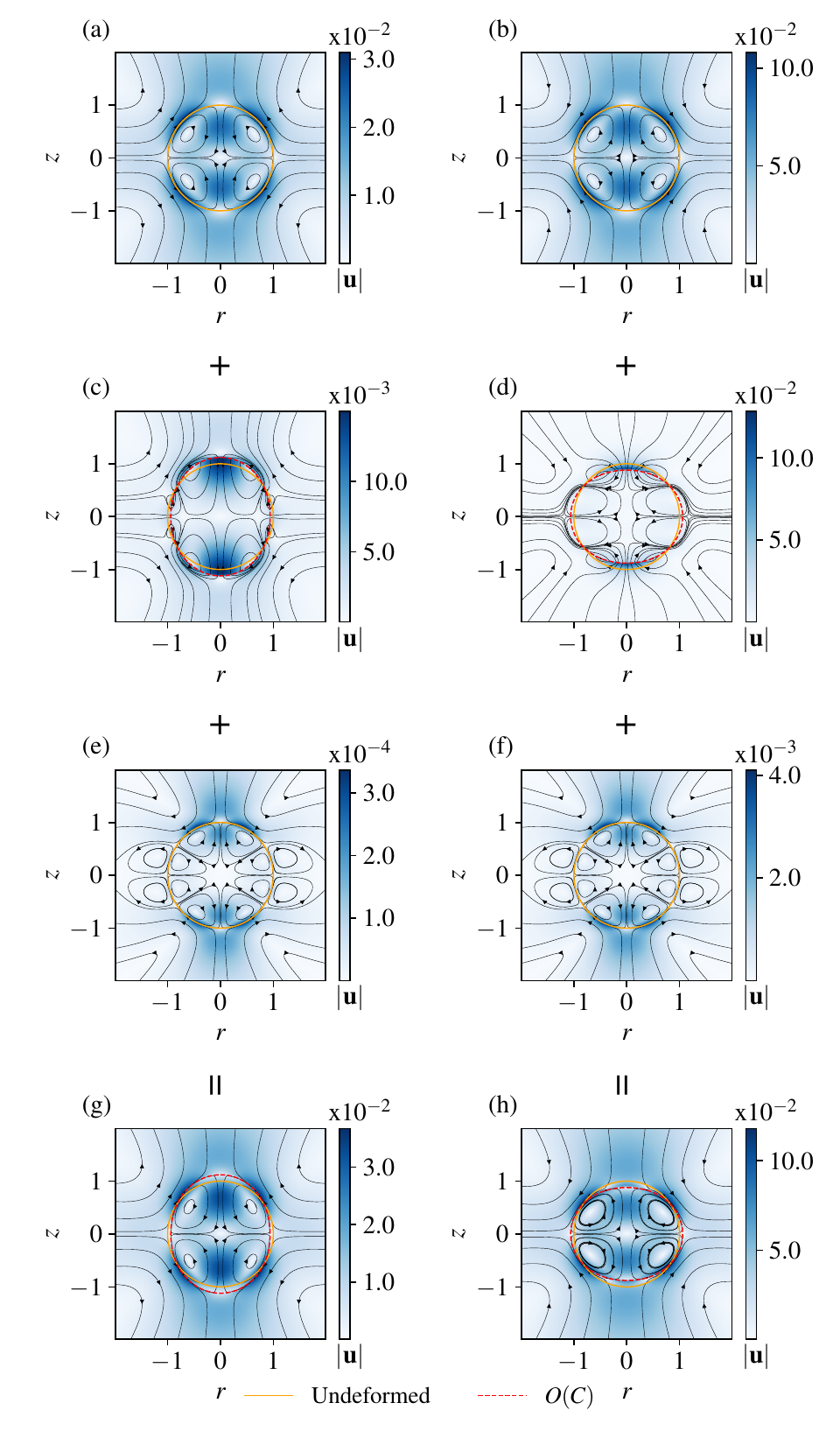}
    \caption{Streamlines for $V_dN_m$ configuration obtained from numerical simulations. $\beta=0.1, \mu_r=1, C=0.2, De=0.2$. The left side shows the contributions for the prolate case, $R=10, Q=0.1$, and the right side for the oblate case, $R=0.5, Q=2$. (a) $O(1)$ to prolate. (b) $O(1)$ to oblate. (c) $O(C)$ to prolate. (d) $O(C)$ to oblate. (e) $O(De)$ to prolate. (f) $O(De)$ to oblate. (g) Final streamlines for prolate. (h) Final streamlines for oblate}
    \label{fig:3_streamlines_analytical}
  \end{minipage}
\end{figure}}

The streamlines shown in \Cref{fig:3_streamlines_analytical} represent the steady-state flow field induced in the drop and surrounding medium by the presence of an electric field. As discussed previously, the viscoelasticity of the drop significantly influences drop deformation; therefore, we plot the streamlines for the $V_dN_m$ configuration. The figures on the left correspond to prolate deformation with parameters $R = 10$ and $Q = 0.1$, while the figures on the right depict oblate deformation with $R = 0.5$ and $Q = 2$. In \Cref{fig:3_streamlines_analytical}(a) and (b), the streamlines for the velocity field at $O(1)$, denoted as $\bm{u}_0$, are presented for the prolate and oblate cases, respectively. Further, \Cref{fig:3_streamlines_analytical}(c) and (d) illustrate the streamlines for the correction of the velocity field at $O(C)$, represented as $C\bm{u}_C$, for both prolate and oblate deformations. Additionally, \Cref{fig:3_streamlines_analytical}(e) and (f) show the streamlines for the correction of the velocity field at $O(De)$, denoted as $De\bm{u}_{De}$ for prolate and oblate cases, respectively. Finally, \Cref{fig:3_streamlines_analytical}(g) and (h) present the streamlines for the total flow field, $\bm{u}_0 + C\bm{u}_C + De\bm{u}_{De}$, for prolate and oblate deformations, respectively. As expected, the streamlines for the flow field $\bm{u}_0$ are tangential to the undeformed drop interface, while the streamlines for the total flow field are tangential to the interface deformed at $O(C)$.  

\Cref{fig:3_streamlines_analytical}(c) illustrates that the $O(C)$ flow field for prolate drops is characterized by elongated recirculation zones aligned with the electric field direction. These zones are complemented by smaller recirculating regions near the equatorial region. In contrast, the $O(C)$ flow field for oblate drops (\Cref{fig:3_streamlines_analytical}(d) exhibits extended recirculation zones perpendicular to the electric field direction, with smaller recirculating regions near the polar region. The $O(De)$ flow field for both prolate and oblate drops (\Cref{fig:3_streamlines_analytical}(e) and (f), respectively) introduces a significant change in the flow pattern. The number of recirculation zones within the drop doubles, and additional recirculation zones emerge near the equatorial region in the ambient medium. Notably, the overall flow direction at all orders remains consistent: from the equator to the pole for prolate drops and from the pole to the equator for oblate drops. Additionally, by examining the color bar labels, it becomes evident that the magnitude of the velocity associated with the $O(De)$ correction is substantially lower than that of the $O(1)$ and $O(C)$ velocities for both prolate and oblate drop shapes. This observation suggests that the influence of viscoelasticity on drop deformation is minimal in the regime of small $De$.

\section{Numerical simulations} \label{sec:numerical}

To analyze large drop deformation, we employ the open-source solver \textsc{Basilisk} \citep{Popinet2003,Popinet2009,Popinet2015}. This solver is based on the finite-volume method, which solves the coupled nonlinear electrohydrodynamic system. One of the key benefits of this solver is its adaptive quadtree spatial discretization which significantly reduces the computational cost. The continuum surface force method is applied in a balanced way to simulate surface-tension-driven flows accurately. The solver incorporates the electric body forces into the Navier-Stokes equation, accounting for charge migration due to both conduction and convection \citep{Lopez-Herrera2011}. The additional stress term arising from viscoelastic studies is also integrated into the Navier-Stokes equation \citep{Lopez-Herrera2019}. To improve the stability of viscoelastic flow simulations, a log conformation approach combined with the time-split scheme is utilized. The interface is tracked using the volume-of-fluid method in which the volume fraction $(c)$ is advected with the fluid using
\begin{equation}
    \pdv{c}{t} + \div(c\vb{u}) = 0 \; .
\end{equation}

The continuity and momentum equations governing the incompressible viscoelastic two-phase flow system are expressed by the one-fluid formulation in dimensionless form as 
\begin{align}
\begin{rcases}
    \div \vb{u} = 0 \; , \\
    Re \left( \pdv{\vb{u}}{t} + \vb{u} \vdot \grad{\vb{u}} \right) =& -\grad{p} + \beta \div \big[ \mu \{ \grad{\vb{u}} + (\grad{\vb{u}})^T \} \big]\\
    &+ \div \boldsymbol{\tau}^{pH} + \frac{1}{C} \vb{F}_s + \vb{F}_e \; ,
\end{rcases}
\end{align}
where $\vb{u}=(u,v)$ is the velocity field. The modified momentum equation incorporates the polymeric stress tensor $(\boldsymbol{\tau}^{pH})$, the surface tension force $(\vb{F}_s)$, and the electric force $(\vb{F}_e)$.

The variation in viscosity is taken as $\mu = c~\mu_r + (1-c)$. The constitutive relation for an Oldroyd-B model is
\begin{equation}
\begin{split}
    De \Bigg[\pdv{\boldsymbol{\tau}^{pH}}{t} + \vb{u} \vdot \grad{ \boldsymbol{\tau}^{pH}} - (\grad{\vb{u}}) \vdot \boldsymbol{\tau}^{pH} &- \boldsymbol{\tau}^{pH} \vdot (\grad{\vb{u}})^T \Bigg]\\
    &+ \boldsymbol{\tau}^{pH} = \left(1-\beta\right) \left(\grad{\vb{u}}+\grad{\vb{u}}^T\right) \; .
\end{split}
\end{equation}

To obtain the electric potential $(\phi)$ and charge density $(q_v)$, we solve the following Poisson equation for the electric potential and the charge transport equation

\begin{align}
\begin{rcases}
    \div \left( \varepsilon \grad{\phi} \right) &= -q_v \; , \\
    Re_E \left[ \pdv{q_v}{t} + \div \left( q_v \vb{u} \right) \right] &= \div \left( \sigma \grad{\phi} \right) \; .
\end{rcases}
\end{align}
For small values of $Re_E$, the volumetric charge density is equivalent to the surface charge density as the charges accumulate quickly at the drop interface. The variation in conductivity and permittivity are taken as $\sigma = c~R + (1-c)$ and $\varepsilon = c~Q + (1-c)$.

\begin{figure}
    \centerline{\includegraphics[width=0.6\textwidth]{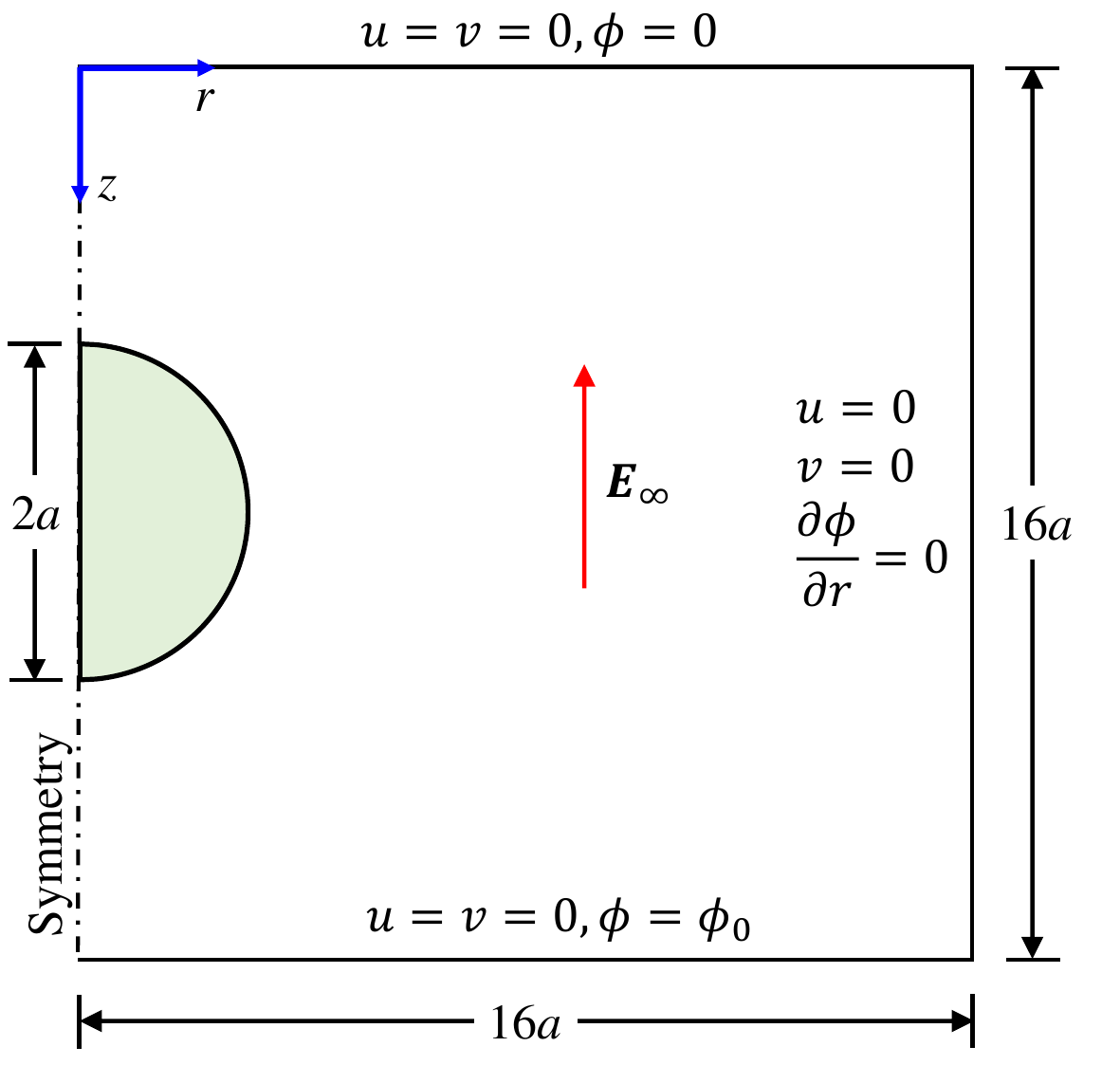}}
    \caption{Schematic representation (not to scale) of the initial configuration and boundary conditions in an axisymmetric coordinate system ($r,z$). The size of the domain is $16a \times 16a$, where $a$ is the radius of the drop. Electric field $(\boldsymbol{E}_\infty)$ is applied along the negative $z$ axis.}
    \label{fig:4_schematic_numerical}
\end{figure}

\begin{figure}
    \centerline{\includegraphics[width=0.72\textwidth]{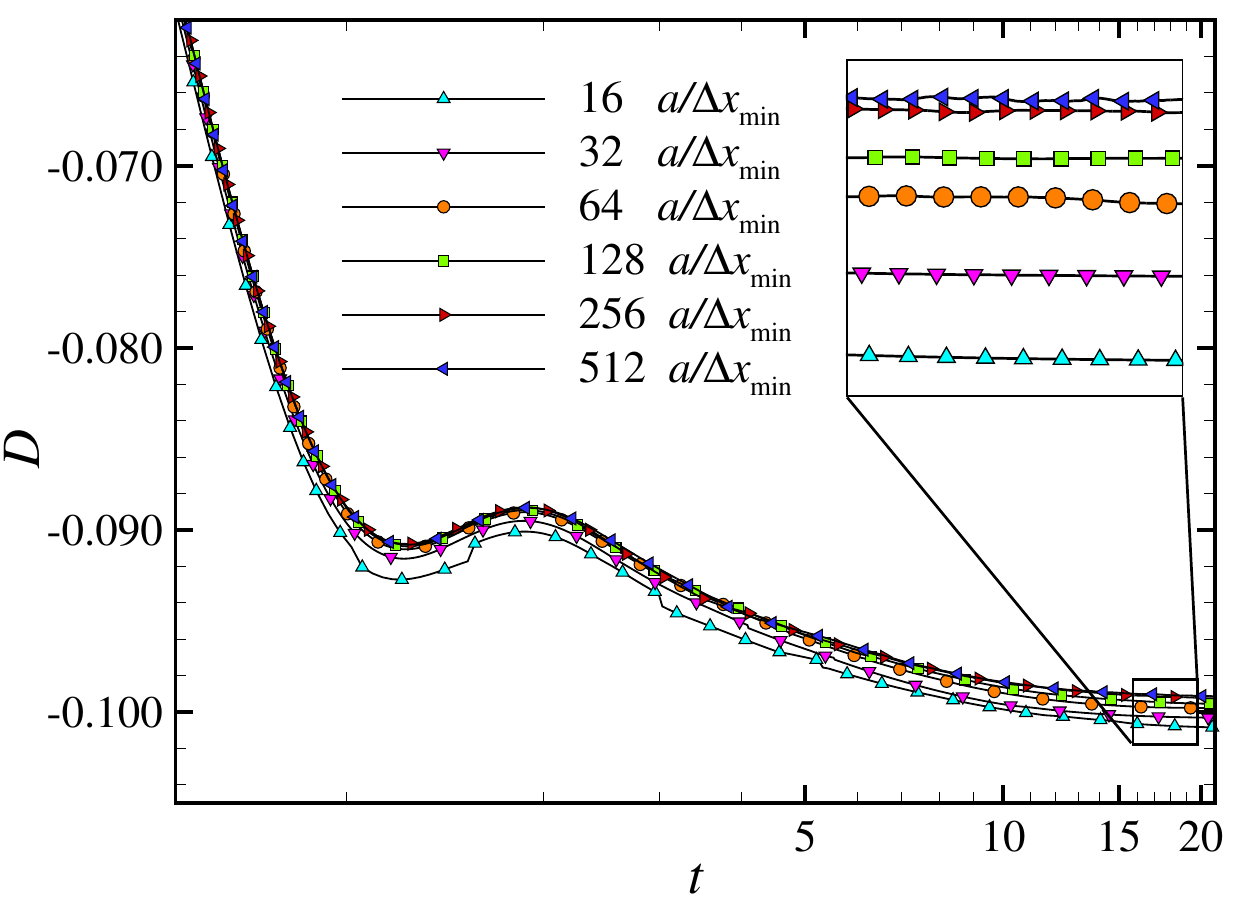}}
    \caption{Temporal evolution of drop deformation ($D$) for different levels of grid refinement. The inset shows the zoomed view for better convergence analysis. The dimensionless parameters for the computations are $R=0.5$, $Q=2$, $De=1$, $C=0.2$, $Re=1$, $\beta=0.1$, $Re_E=0.01$ and $\mu_r=1$.}
    \label{fig:5_grid_converge}
\end{figure}

\Cref{fig:4_schematic_numerical} shows a schematic of the computational domain with the boundary conditions. An external electric field is applied along the axial direction by applying an electric potential at the bottom boundary and grounding the top boundary. Symmetry boundary conditions are imposed on the axis of symmetry, while no-slip and impervious velocity conditions are imposed on other boundaries. Simulations are performed on a two-dimensional axisymmetric domain of size $16a \times 16a$, where $a$ is the radius of the drop. A grid convergence study is performed to ensure the results are independent of the grid resolution. The adaptive meshing capability of the solver allows us to use a very fine mesh near the interface and a relatively coarser mesh away from the drop. This results in less computational time than a uniformly refined grid. The entire interface is resolved using $\Delta x_{min}$, and the grid is progressively coarsened away from the drop. The deformation of the drop is characterized by Taylor's deformation parameter, $D$, where $D = (L - B)/(L + B)$. Here, $L = r_s(0)$ represents the dimension of the deformed drop in the direction of the electric field, and $B = r_s(\pi/2)$ represents the dimension perpendicular to the electric field. The temporal evolution of the drop deformation ($D$) of a viscoelastic drop suspended in a Newtonian fluid is plotted in \Cref{fig:5_grid_converge}. It can be observed from the inset that the curves corresponding to $a/\Delta x_{min}$ = 256 ($\text{level} = 12$) and 512 ($\text{level} = 13$) almost overlap, and hence a further grid refinement would have a negligible effect. Therefore, a refinement level of 12 has been selected for the remainder of the study.

\subsection{Comparison between analytical and numerical results}

\begin{figure}[h]
    \centerline{\includegraphics[width=\textwidth]{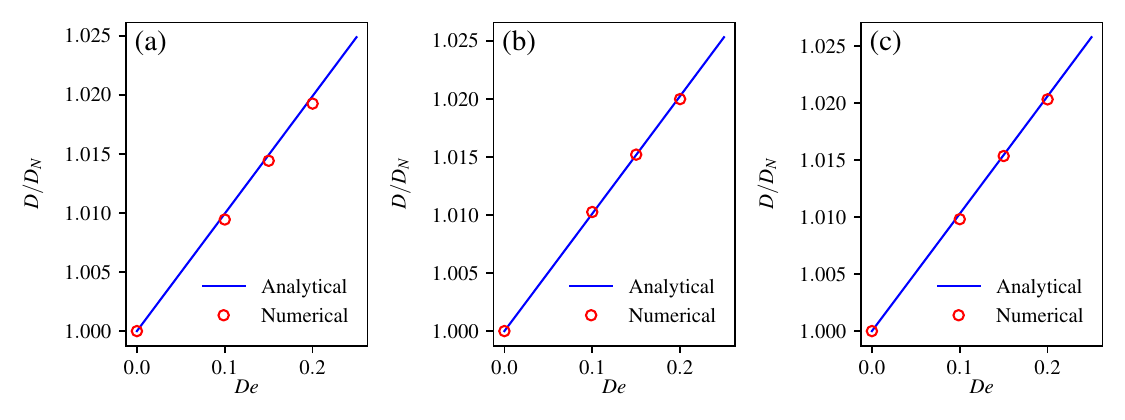}}
    \caption{Drop deformation ($D$) variation with $De$ for three different (a) $C = 0.10$, (b) $C = 0.15$, and (c) $C = 0.20$. The dimensionless parameters for the computations are $R=0.5$, $Q=2$, $De=1$, $C=0.2$, $Re=1$, $\beta=0.1$, $Re_E=0.01$ and $\mu_r=1$.}
    \label{fig:6_ana_num_compare}
\end{figure}

We perform a comparative analysis between the results of the current analytical study and full-scale computational simulations, as illustrated in \Cref{fig:6_ana_num_compare}. The analytical solution is derived for an unbounded domain. We plot the deformation characteristics of a viscoelastic drop, normalized against the deformation of a Newtonian drop, as a function of the Deborah number, $De$, which is a critical parameter in characterizing the viscoelastic properties of the fluid. The results for electric capillary numbers, $C$, of 0.1, 0.15, and 0.2 are shown in \Cref{fig:6_ana_num_compare}(a), (b), and (c), respectively. The solid blue lines represent the predictions from the asymptotic analysis,
while the red markers indicate the results from numerical simulations. There is a clear agreement between the analytical and numerical solutions. \Cref{fig:7_streamlines_numerical}(a) and (b) present the streamlines generated from numerical simulations for prolate and oblate drops, respectively. These simulations employ identical parameters as those used in the analytical calculations presented in \Cref{fig:3_streamlines_analytical}. There is an excellent agreement between the analytical and numerical results.

\textbf{\begin{figure}
    \centering
    \includegraphics[width=\textwidth]{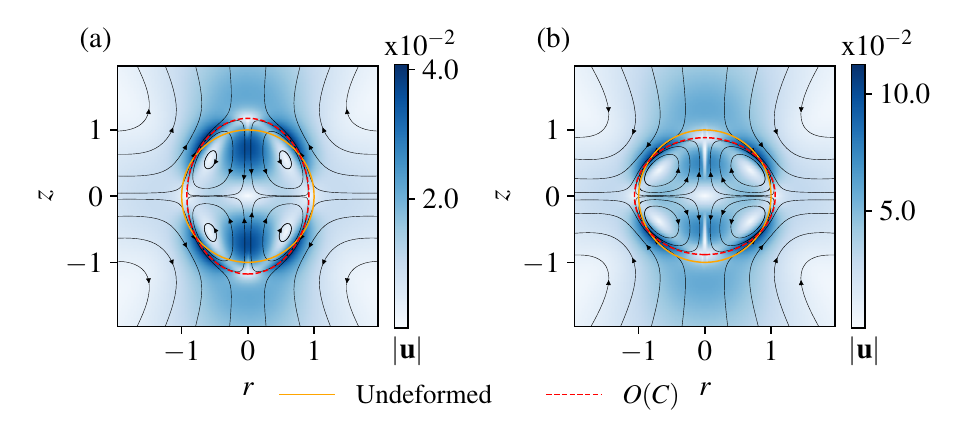}
    \caption{Streamlines for $V_dN_m$ configuration. $\beta=0.1, \mu_r=1, C=0.2, De=0.2$. (a) prolate case, $R=10, Q=0.1$ (b) oblate case, $R=0.5, Q=2$.}
    \label{fig:7_streamlines_numerical}
\end{figure}}

\subsection{Large deformation}

In this section, we simulate large deformation cases for two configurations: a viscoelastic drop suspended in a Newtonian fluid ($V_dN_m$) and a Newtonian fluid suspended in a viscoelastic medium ($N_dV_m$). The application of a uniform electric field induces deformation in the drop. Due to the presence of the viscoelastic fluid in the medium, we observe different dynamics compared to those seen with Newtonian fluids. For all the studies, we keep the following parameters as constants: $R=10$, $Q=0.1$, $Re=1$, $Re_E = 0.01$, $\beta=0.1$, $\rho_r=1$, and $\mu_r=1$. The governing dimensionless parameters for these studies include $R$, $Q$, $C$, and $De$. We will first present the transient dynamics of drop deformation, followed by the steady-state dynamics.

\subsubsection{Transient drop deformation}

\begin{figure}[h]
    \centering
    \begin{subfigure}[b]{0.49\textwidth}
    \includegraphics[width=\textwidth]{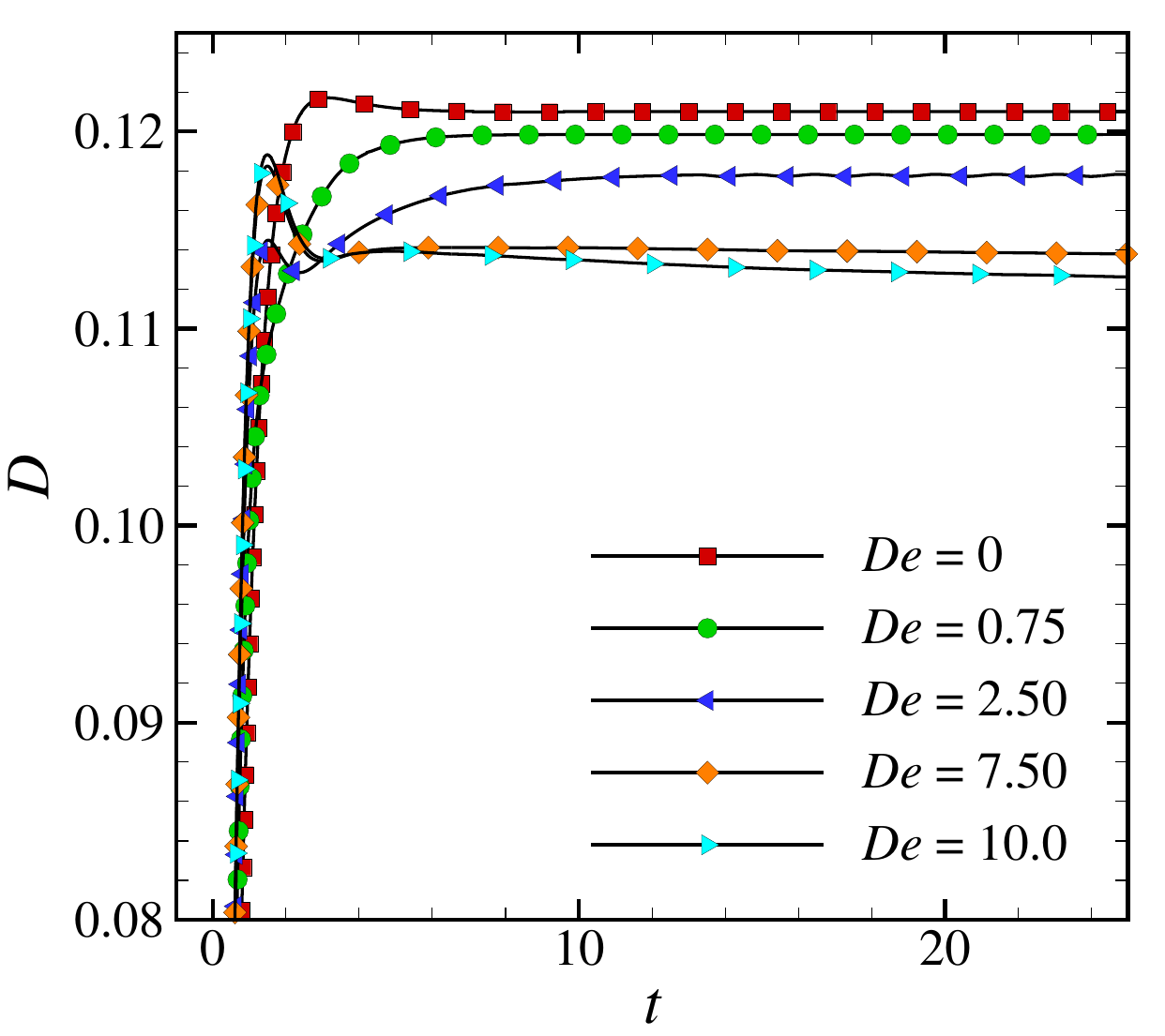}
    \caption{$V_dN_m$} 
    \end{subfigure}
    \hfill   
    \begin{subfigure}[b]{0.49\textwidth}
    \includegraphics[width=\textwidth]{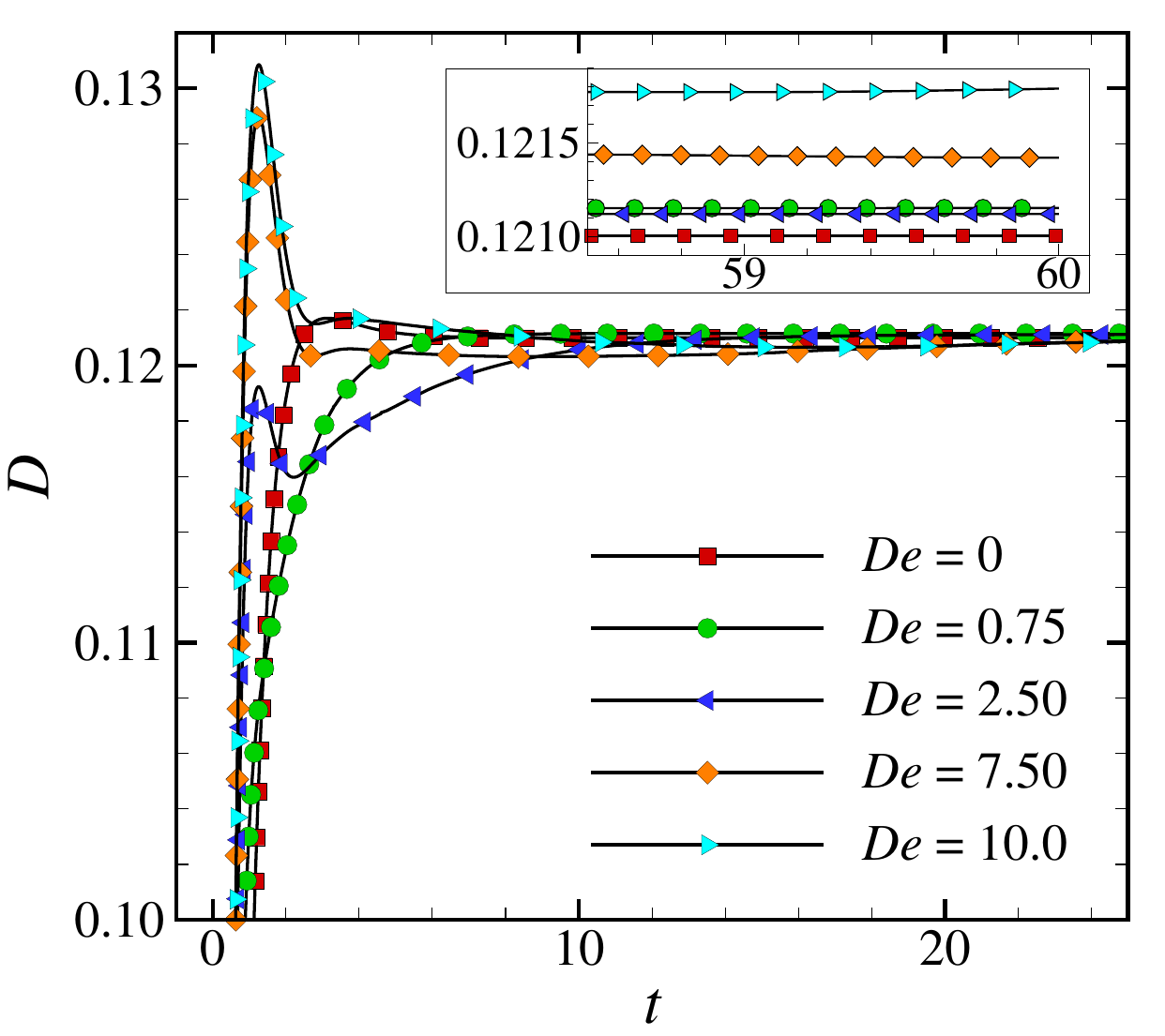}
    \caption{$N_dV_m$}
    \end{subfigure}             
    \caption{Effect of $De$ on deformation of drop for $R=10$, $Q=0.1$ and $C=0.2$. The inset for the $N_dV_m$ configuration shows that at steady-state, the deformation increases with an increase in $De$.}
    \label{fig:8_transient}
\end{figure}

\Cref{fig:8_transient} shows the drop deformation as a function of time for $C=0.2$ and varying $De$. We also plot the deformation history for a Newtonian drop represented by $De=0$. In the $V_dN_m$ configuration, the steady-state deformation $D$ decreases as $De$ increases. This is due to the increased viscoelastic stresses, which hinder drop deformation. Conversely, in the $N_dV_m$ configuration, viscoelastic stresses facilitate drop deformation, resulting in an increase in $D$ with an increase in $De$, although there is a very slight deviation from the Newtonian case.

We also observe overshoots and oscillations in the transient deformation for both configurations. These phenomena are attributed to the relaxation time associated with the Oldroyd-B model, which introduces a finite time delay for the development of viscoelastic stresses. Similar variations in temporal deformations have been reported in previous studies \citep{Yue2005,Aggarwal2007}. The differences in steady-state deformation with varying $De$ can be traced back to the nonlinear nature of the Oldroyd-B constitutive equation, which leads to a finite first normal stress difference. The deformation is particularly pronounced in the $V_dN_m$ configuration, while changes in deformation for the $N_dV_m$ configuration are minimal. 

\subsubsection{Steady state drop deformation}

\begin{figure}[h]
    \centering
    \begin{subfigure}[b]{0.48\textwidth}
    \includegraphics[width=\textwidth]{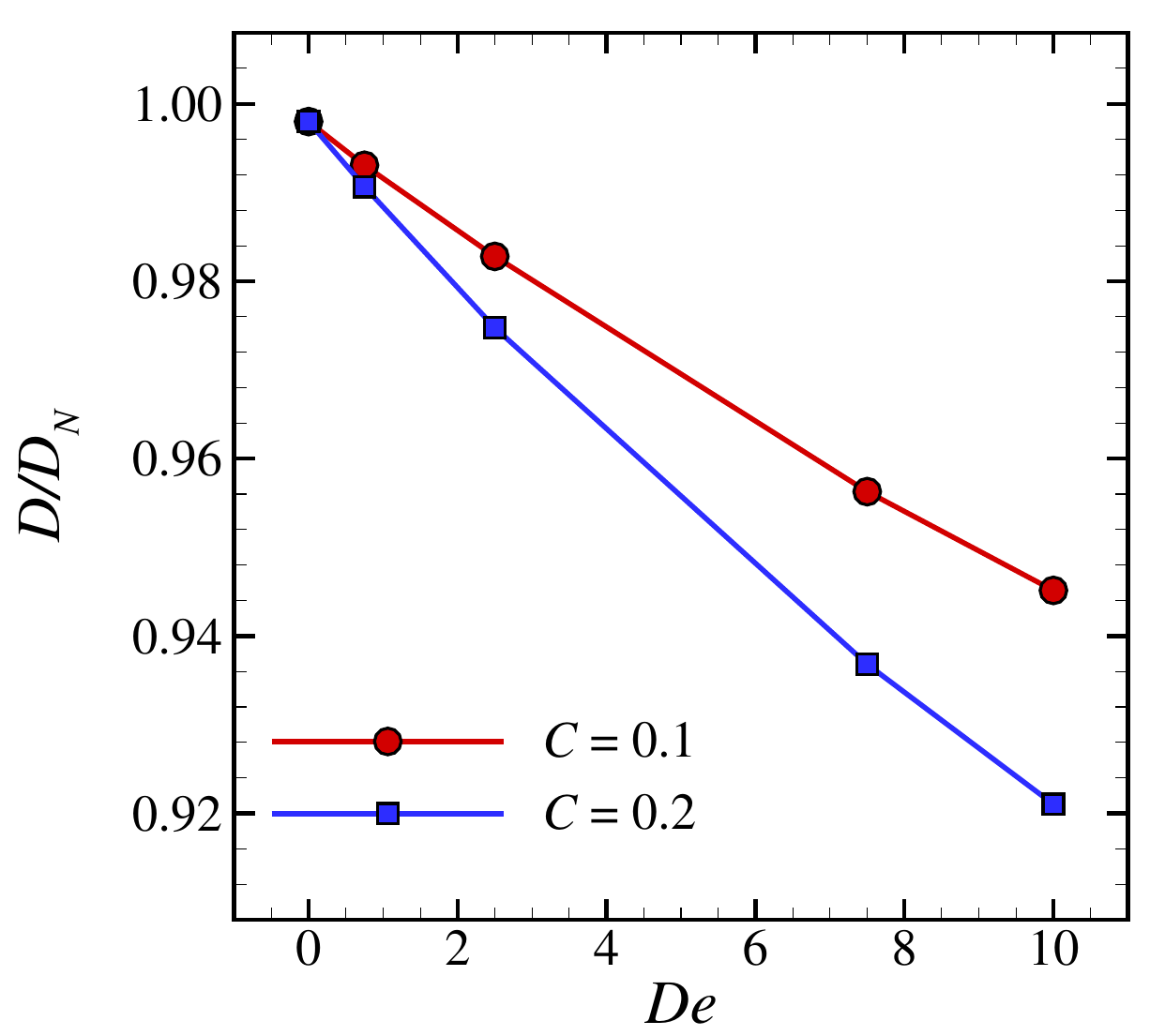}
    \caption{$V_dN_m$}
    \end{subfigure}
    \hfill   
    \begin{subfigure}[b]{0.48\textwidth}
    \includegraphics[width=\textwidth]{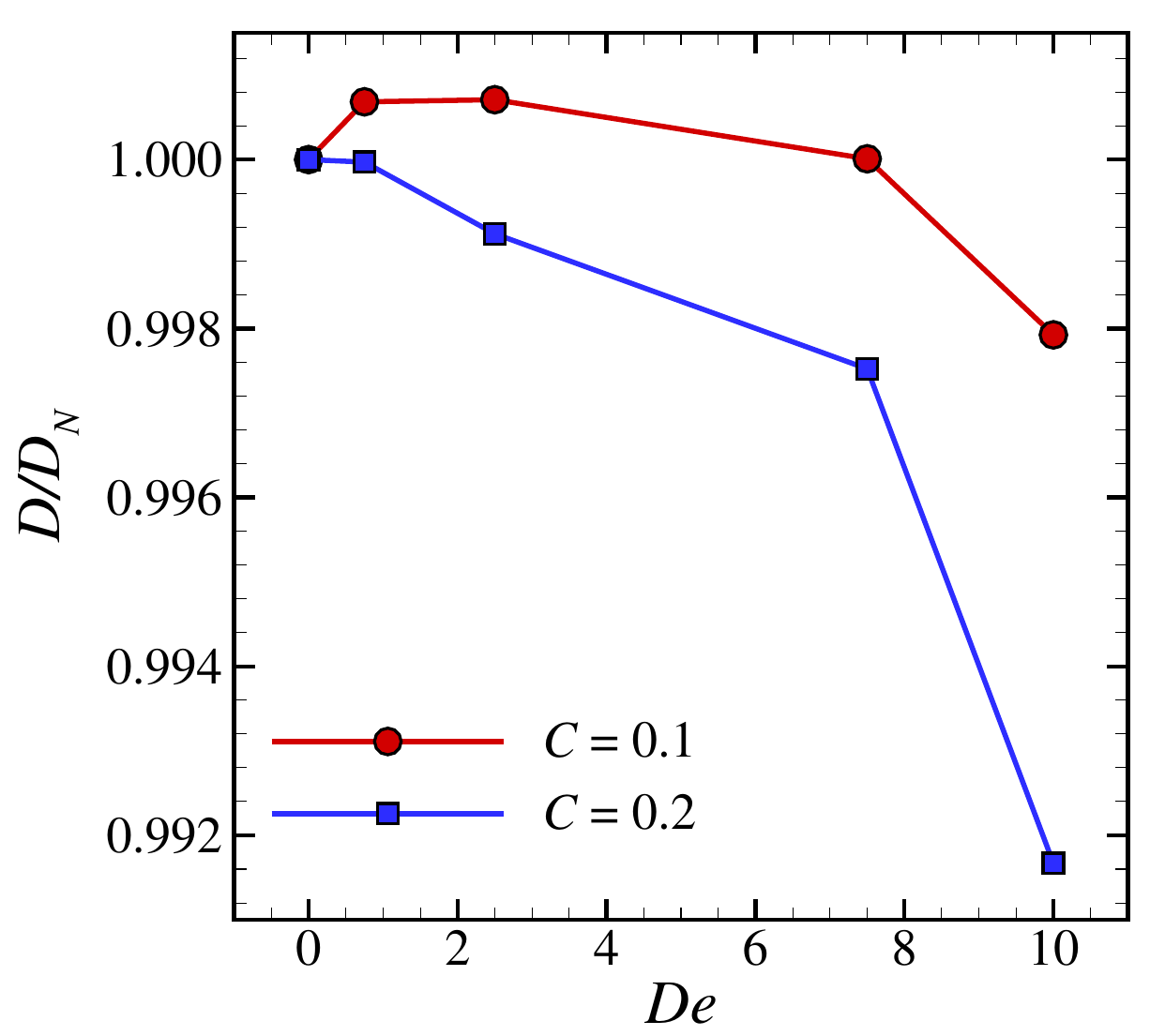}
    \caption{$N_dV_m$}
    \end{subfigure}             
    \caption{Steady-state drop deformation variation $D/D_N$ with $De$ for $C = 0.1$ and $0.2$. The effect of viscoelasticity is more prominent in the $V_dN_m$ system.}
    \label{fig:9_steady}
\end{figure}

The steady-state drop deformation values are normalized by the Newtonian drop deformation $D_N$ at $De=0$. In \Cref{fig:9_steady}, the plot of $D/D_N$ as a function of $De$ is presented for two values of $C$. For the $V_dN_m$ configuration, the deformation decreases consistently as $De$ increases. Similarly, the deformation decreases for the $N_dV_m$ configuration also, although it shows a non-monotonic behavior. This indicates that the deformation of the drop continuously decreases with $De$, and the deformation is greater when the drop is viscoelastic. This was also the observation from the small deformation theory. The stress fields surrounding the drop are examined to understand this behavior further.

For the electrical properties with $R = 10$ and $Q = 0.1$, the normal electric stresses at the poles lead to an increase in the length $L$ of the drop in the direction of the electric field. In contrast, at the equator, these normal electric stresses reduce the width $B$ of the drop. As expected, this results in an overall prolate deformation of the drop. Since the electric stresses do not depend directly on $De$, we do not show their variation and focus instead on the changes in viscous and viscoelastic stresses.

We evaluate the stresses exerted on the circumference of the drop interface as a function of the angular position, represented by $\theta$, relative to the direction of the electric field. The angular position is normalized ($\theta/\pi$) to a
\newpage
\begin{figure}[ht]
    \centering
    \begin{subfigure}[b]{0.49\textwidth}
    \includegraphics[width=\textwidth]{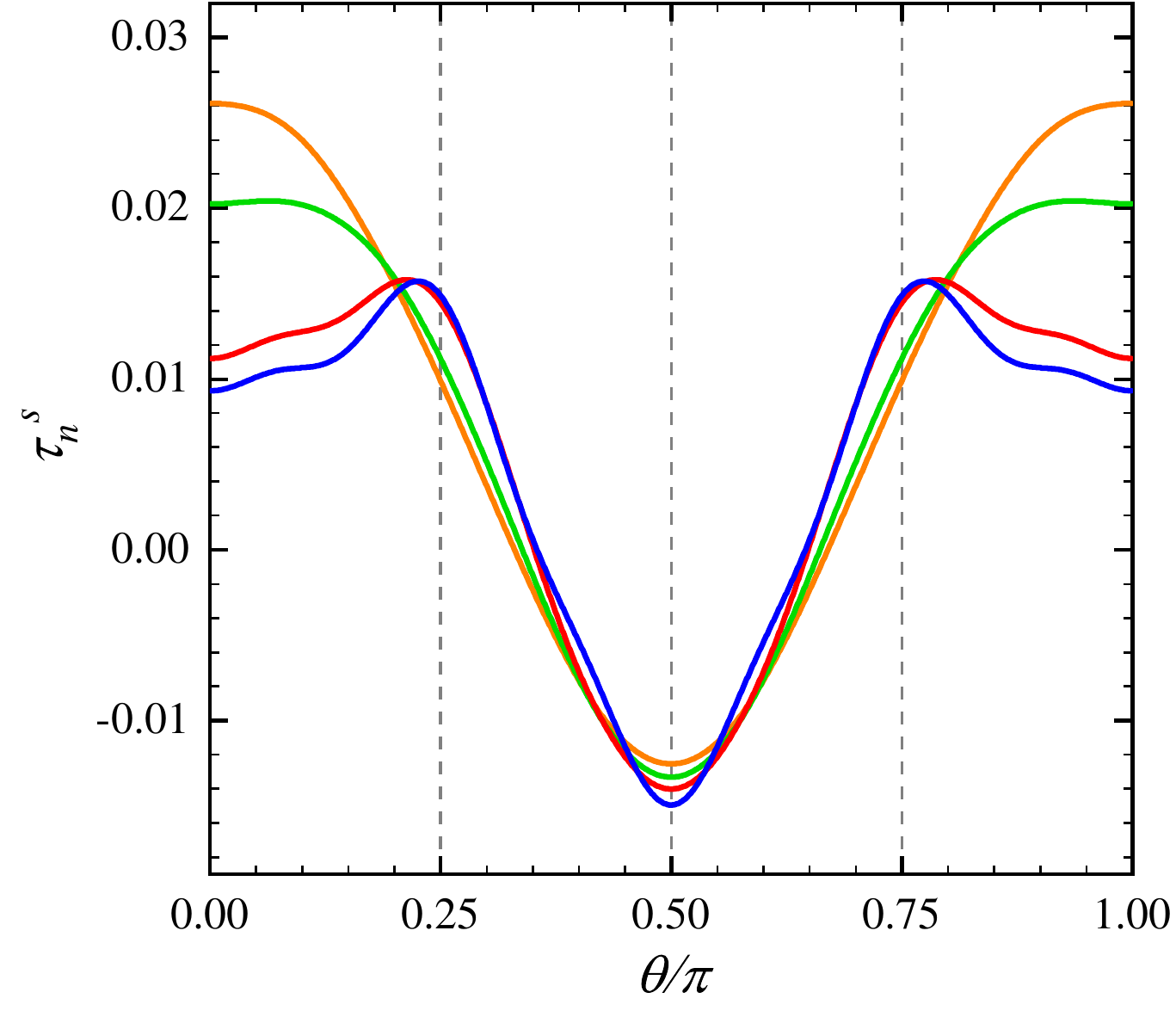}
    \caption{}
    \end{subfigure}
    \hfill   
    \begin{subfigure}[b]{0.49\textwidth}
    \includegraphics[width=\textwidth]{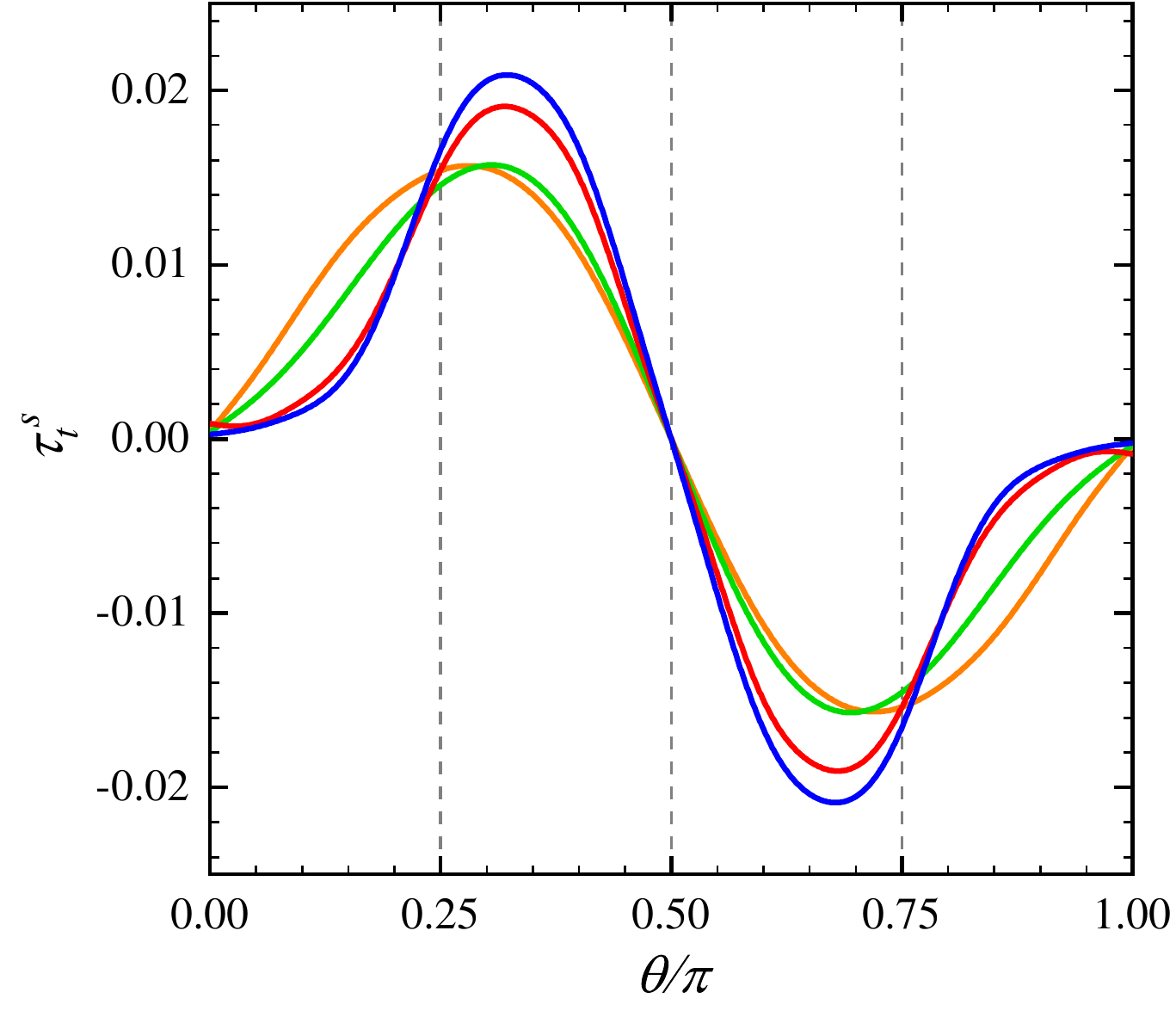}
    \caption{}
    \end{subfigure}
    \hfill
    \begin{subfigure}[b]{0.49\textwidth}
    \includegraphics[width=\textwidth]{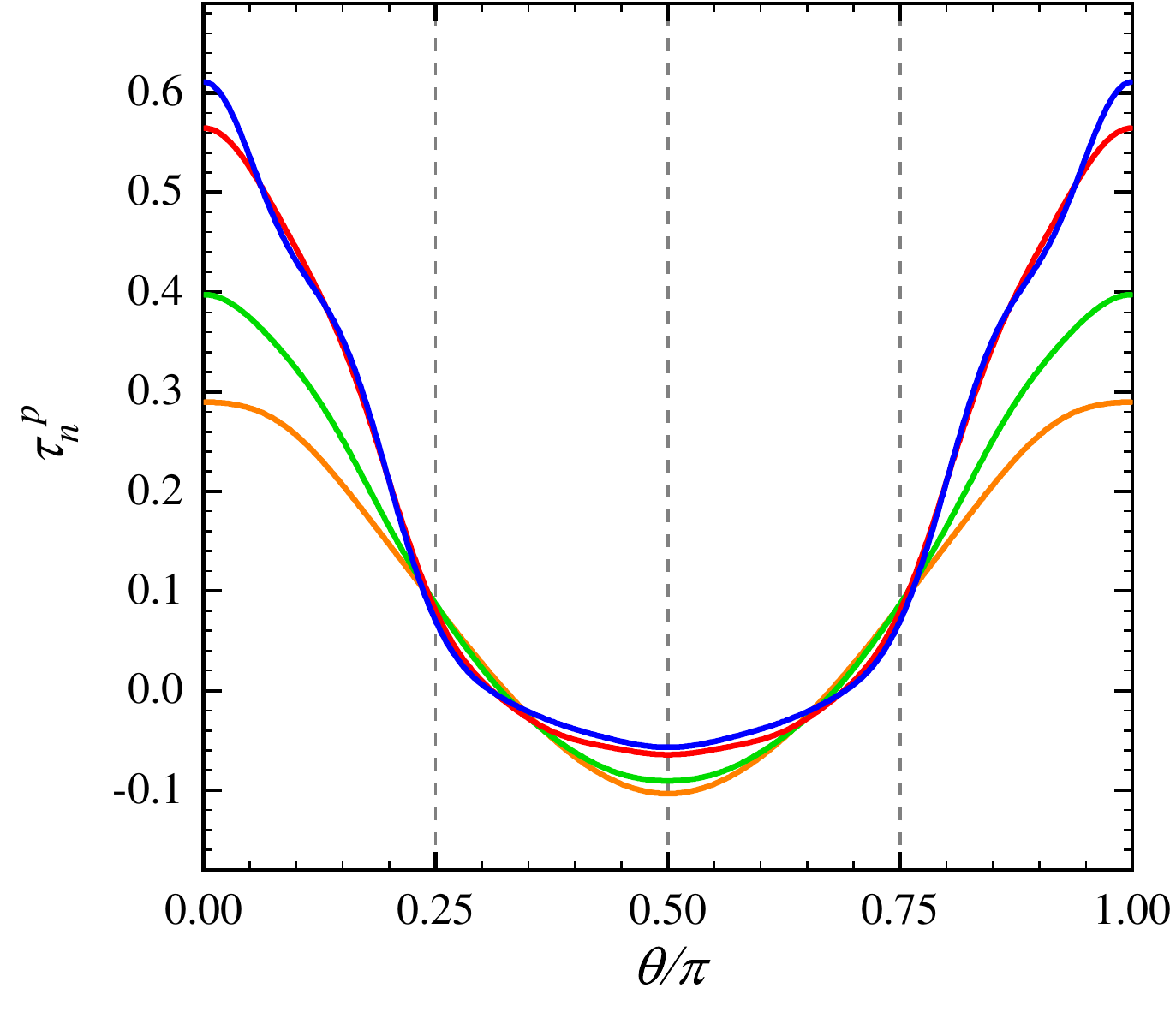}
    \caption{}
    \end{subfigure}
    \hfill   
    \begin{subfigure}[b]{0.49\textwidth}
    \includegraphics[width=\textwidth]{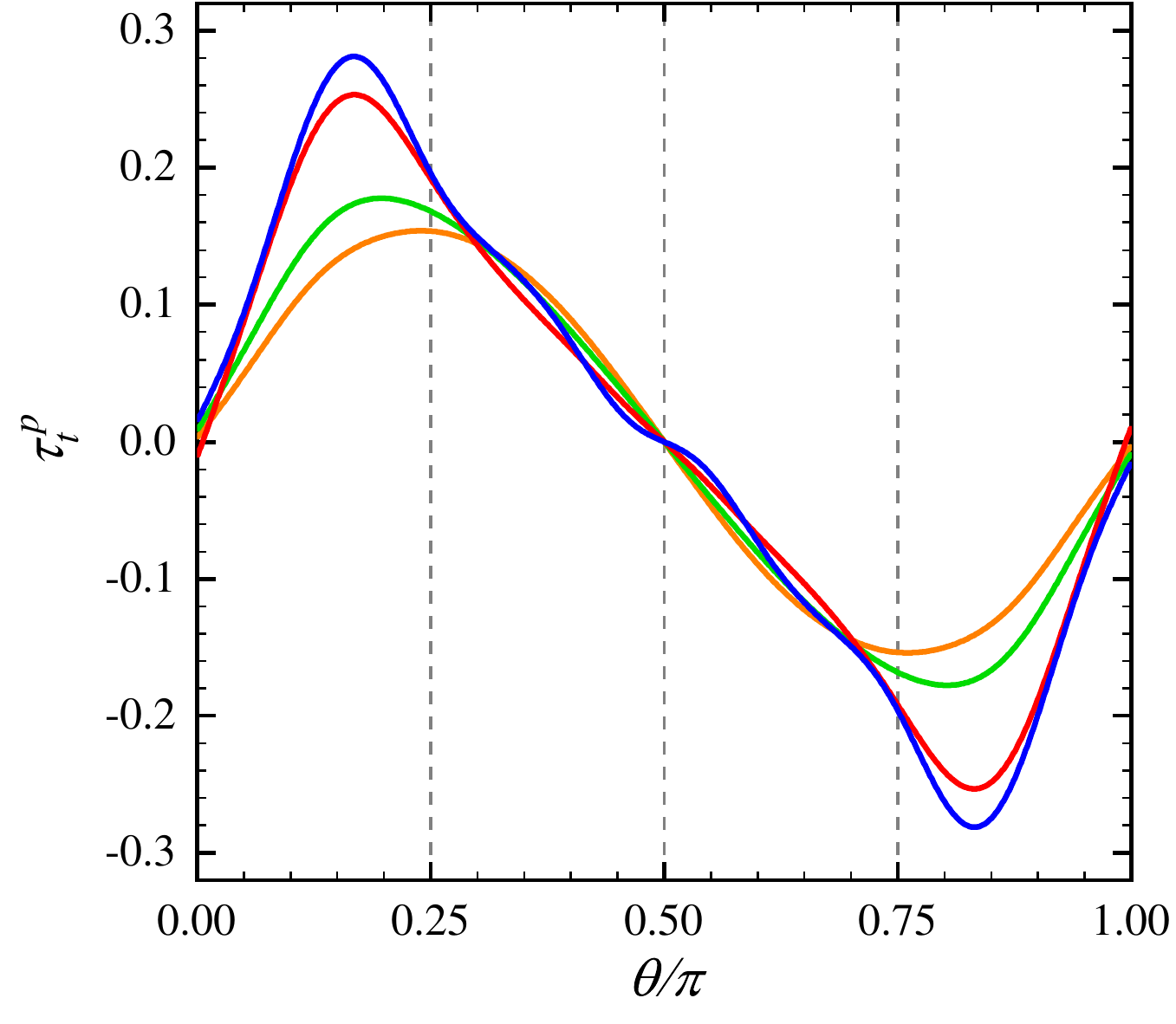}
    \caption{}
    \end{subfigure}
    \caption{Various stresses along the inner edge of the interface for $V_dN_m$ configuration at $C=0.1$ with $n$ being the inward normal to the interface. (a) The viscous normal stress $\tau^s_n$ and (b) the viscous tangential stress $\tau^s_t$. (c) The polymer normal stress $\tau^p_n$ and (d) the polymer tangential stress $\tau^p_t$ due to the solvent in the Oldroyd-B fluid.}
    \label{fig:10_stress}
\end{figure}
\noindent range between 0 and 1, where 0 and 1 represent the poles, while a value of 0.5 indicates the equator. The orange line denotes $De=0.75$, green line denotes $De=2.5$, red line denotes $De=7.5$, and blue line denotes $De=10$. First, we analyze the stresses for a $V_dN_m$ configuration. In \Cref{fig:10_stress}(a) and \Cref{fig:10_stress}(c), we plot the normal components of the viscous stress ($\tau^s_n$) and viscoelastic stress ($\tau^p_n$) along the inner edge of the drop. The viscous normal stress decreases with increasing $De$ at the poles while it remains relatively constant at the equator. In contrast, the viscoelastic normal stress increases as $De$ increases. Notably, these viscoelastic normal stresses are greatest near the drop poles, exerting an inward pull that reduces prolate extension $L$. Furthermore, a slight reduction in viscoelastic normal stress at the equator is observed as $De$ increases, which exerts an inward push that decreases $B$. It is important to note that the magnitude of the viscous normal stress is significantly lower than that of the viscoelastic normal stress, indicating that they do not contribute substantially to the deformation dynamics.

The tangential stresses along the drop interface are shown in \Cref{fig:10_stress}(b) and \Cref{fig:10_stress}(d). The viscous tangential stresses ($\tau^s_t$) decrease at the poles as $De$ increases, whereas the viscoelastic tangential stress ($\tau^p_t$) at the poles remains essentially unchanged with increasing $De$. Both the viscous and the viscoelastic tangential stresses are tensile near the equator of the drop and increase with $De$. Again, the magnitude of the viscous tangential stresses is relatively small compared to the viscoelastic stresses, highlighting that the effects of viscoelasticity dominate. The drop interface experiences a net inward pull due to the hoop stress effect generated by the viscoelastic tensile stresses. This results in high pressure at the drop equator, which pushes the drop outward, thereby increasing $B$. In conclusion, the viscoelastic normal stresses at the poles decrease $L$, while the viscoelastic tangential stresses at the equator increase $B$. Collectively, these effects lead to an overall reduction in drop deformation as $De$ increases.

\begin{figure}[ht]
    \centering
    \begin{subfigure}[b]{0.49\textwidth}
    \includegraphics[width=\textwidth]{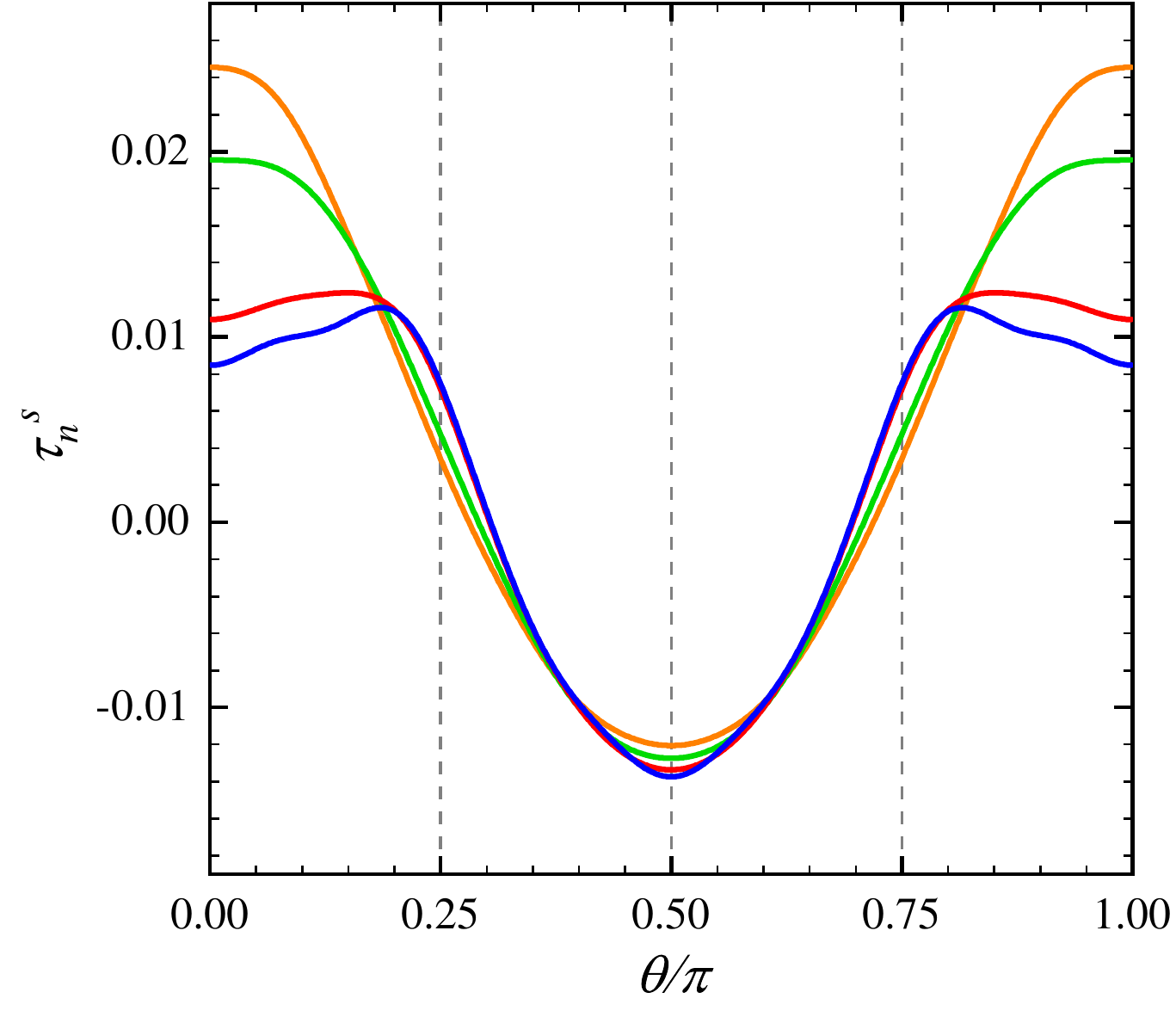}
    \caption{} 
    \end{subfigure}
    \hfill   
    \begin{subfigure}[b]{0.49\textwidth}
    \includegraphics[width=\textwidth]{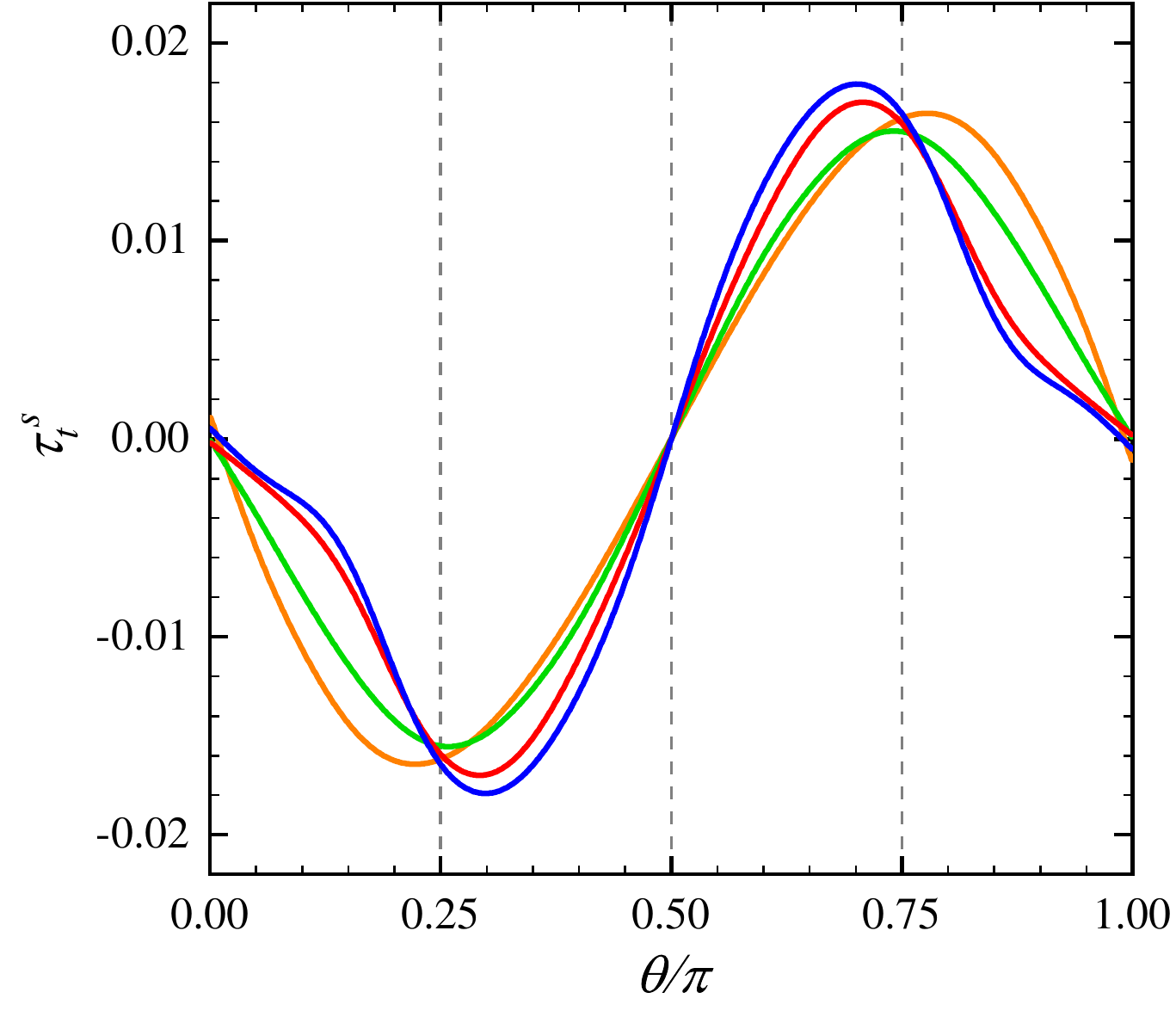}
    \caption{}
    \end{subfigure}
    \hfill
    \begin{subfigure}[b]{0.49\textwidth}
    \includegraphics[width=\textwidth]{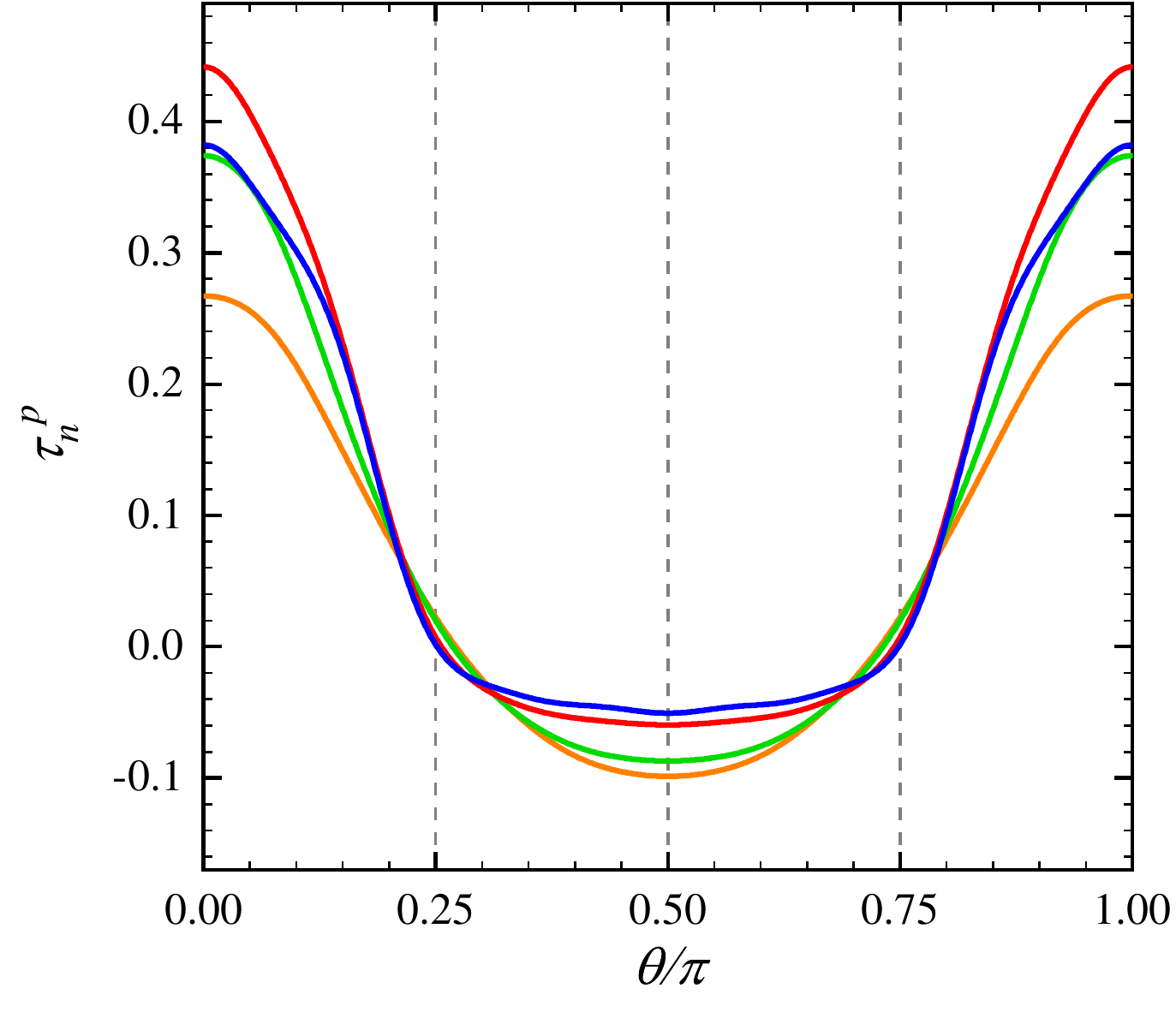}
    \caption{}
    \end{subfigure}
    \hfill   
    \begin{subfigure}[b]{0.49\textwidth}
    \includegraphics[width=\textwidth]{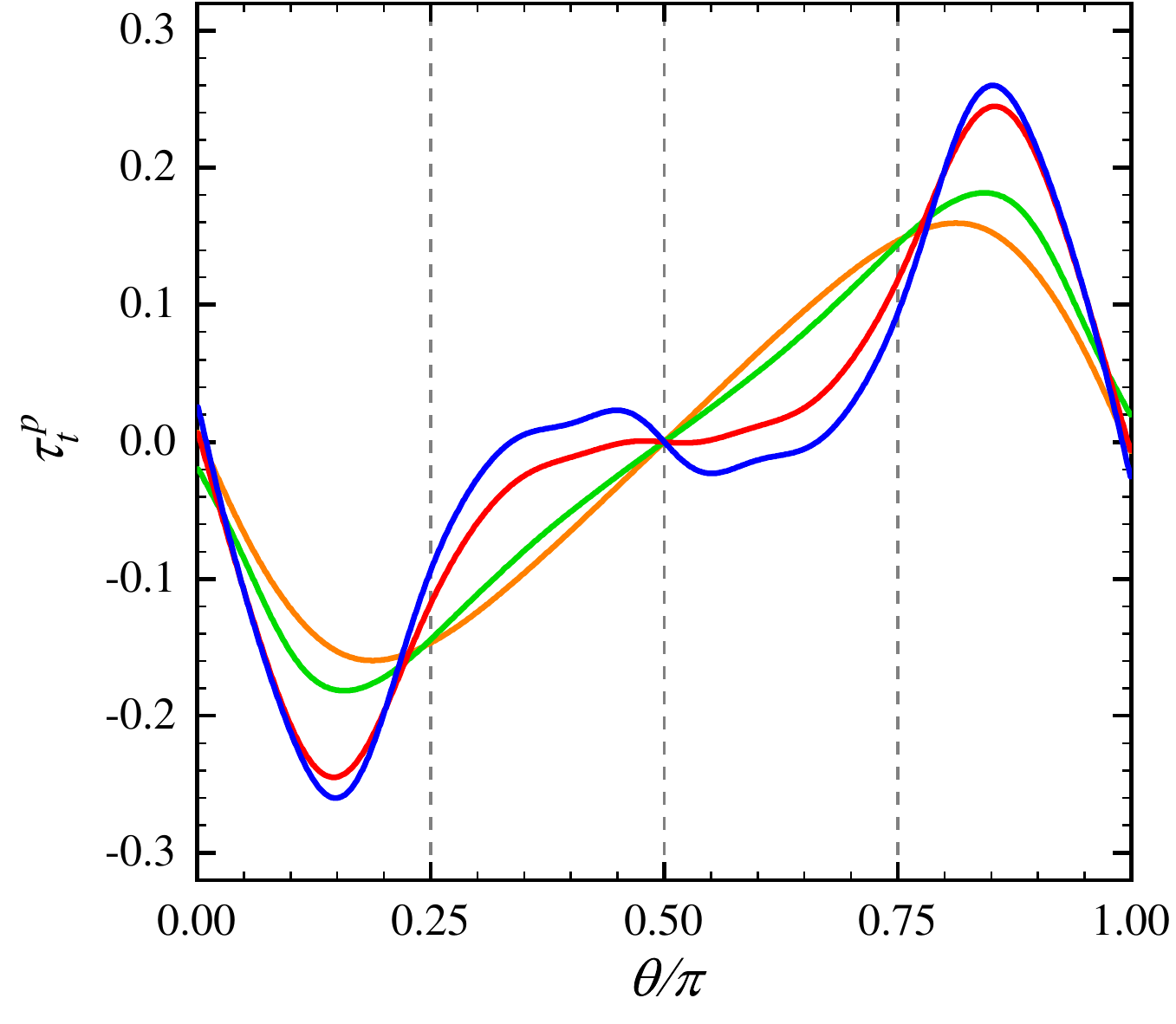}
    \caption{}
    \end{subfigure}
    \caption{Various stresses along the inner edge of the interface for $N_dV_m$ configuration at $C=0.1$ with $n$ being the outward normal to the interface. (a) The viscous normal stress $\tau^s_n$ and (b) the viscous tangential stress $\tau^s_t$. (c) The polymer normal stress $\tau^p_n$ and (d) the polymer tangential stress $\tau^p_t$ due to the solvent in the Oldroyd-B fluid.}
    \label{fig:11_stress}
\end{figure}

Now, we shift our focus to the $N_dV_m$ configuration. In this study, we observe a general decrease in drop deformation with increasing $De$, along with some non-monotonic behavior. In \Cref{fig:11_stress}(a) and \Cref{fig:11_stress}(c), we plot the normal components of the viscous ($\tau^s_n$) and viscoelastic ($\tau^p_n$) stresses along the outer edge of the drop interface for various $De$ values. The viscous normal stress is approximately ten times lower than the viscoelastic normal stress. At the poles, the viscous normal stress decreases as $De$ increases, while near the equator of the drop, these stresses exhibit only slight variation. As elasticity increases, the magnitude of the viscoelastic normal stress at the poles rises significantly, whereas at the equator, it decreases.

The viscoelastic normal stress at the poles shows a weak non-monotonic behavior, initially increasing and then decreasing. This leads to a change in $L$, which, although relatively small, contributes to the overall non-monotonic nature of the deformation. It is noteworthy that this non-monotonic behavior of viscoelastic normal stress at the poles is not observed at higher values of $C$ (not shown here), leading to a decrease in deformation (see \Cref{fig:9_steady}(b)). The tangential components of the viscous ($\tau^s_t$) and viscoelastic ($\tau^p_t$) stresses
\newpage
\noindent are shown in \Cref{fig:11_stress}(b) and \Cref{fig:11_stress}(d). At the poles, both the viscous and viscoelastic tangential stresses do not exhibit significant variation with respect to $De$. However, near the equator of the drop, the viscoelastic tangential stress increases while the viscous tangential stress decreases as $De$ increases. Given that the magnitude of the viscous tangential stresses is relatively small compared to the viscoelastic tangential stresses, the drop interface experiences a net inward pull, which generates high pressure at the equator of the drop. This pressure pushes the drop outward, thereby increasing $B$. In summary, we observe a non-monotonic change in $L$ due to the viscoelastic normal stresses, while $B$ increases, characterized by an increase in viscoelastic tangential stress.

\begin{figure}[h]
    \centering
    \begin{subfigure}[b]{0.49\textwidth}
    \includegraphics[width=\textwidth]{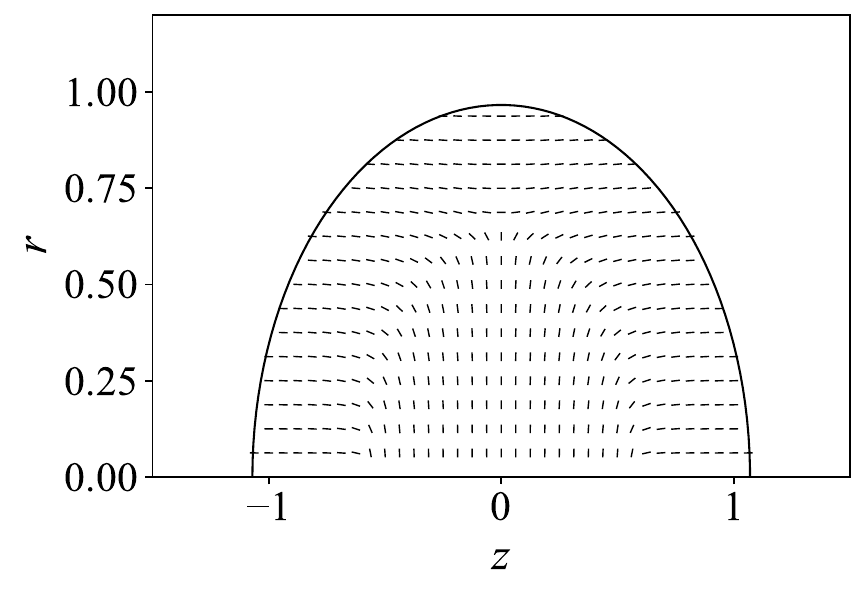}
    \caption{$V_dN_m$} 
    \end{subfigure}
    \hfill   
    \begin{subfigure}[b]{0.48\textwidth}
    \includegraphics[width=\textwidth]{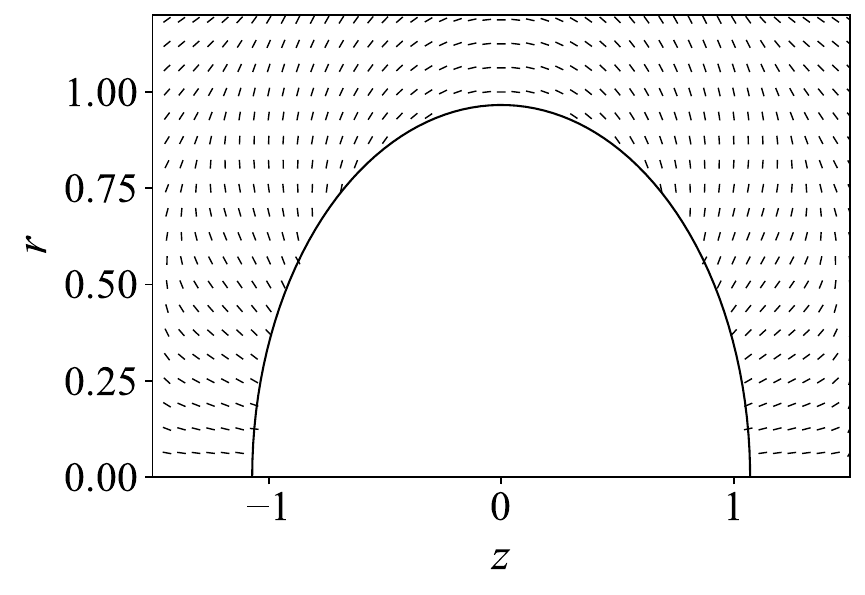}
    \caption{$N_dV_m$}
    \end{subfigure}             
    \caption{Orientation of dominant polymer molecules for the configurations (a) $V_dN_m$ and (b) $N_dV_m$ at $C=0.1$ and $De=0.75$.}
    \label{fig:12_orientation}
\end{figure}

In \Cref{fig:12_orientation}, we plot the primary eigenvalue of the conformation tensor $(=(\lambda/\mu)\boldsymbol{\tau}^{p}+\mathbf{I})$. The conformation tensor indicates the polymer orientation and stretching at each node point throughout the material and is crucial for understanding the behavior of polymer molecules in a viscoelastic. The orientation of the polymer molecules shown in \Cref{fig:12_orientation} aligns with our earlier observations regarding the trends in viscoelastic tangential and normal stresses at the interface. We observe that in regions near the poles of the drop, the polymer molecules orient themselves perpendicularly, suggesting a significant influence on the normal stress behavior in those areas. In contrast, around the equator, these molecules tend to be more aligned tangentially to the interface.

\section{Conclusions} \label{sec:conclusions}

In this study, we investigated the effect of viscoelasticity on drop deformation under the influence of an electric field. Previous studies on viscoelasticity related to drop deformation have only focused on shear and extensional flows, highlighting a significant gap in understanding how these properties influence deformation under electric fields. Our investigation includes a comprehensive analytical and numerical analysis of drop behavior when subjected to a steady and uniform electric field. This study considers two configurations: a viscoelastic drop suspended in a Newtonian fluid and a Newtonian drop suspended in a viscoelastic matrix. The primary focus of our investigation is the effect of Deborah number. This dimensionless parameter quantifies the relative importance of fluid relaxation time to the time scale of the flow. To model these interactions, we employ the Oldroyd-B constitutive model, which effectively captures the viscoelastic behavior of the fluids. It is essential to note that while we examine the effects of viscoelasticity, all other properties, including the electrical properties of the fluids, are maintained constant to isolate the influence of viscoelastic properties.

To investigate the drop dynamics under small deformations, we utilize two critical parameters as perturbation factors: the Deborah number, $De$, and the electric capillary number, $C$, which measures the relative effect of electrical forces to surface tension forces. We utilize the method of domain perturbations to evaluate the boundary conditions at the deformed interface of the drop as it responds to the electric field. We derive approximate closed-form expressions for drop deformation up to orders $O(C)$, $O(C^2)$, and $O(CDe)$. This provides a clearer insight into how these factors influence the dynamics of drop deformation. Our analysis reveals that the viscoelastic properties of the drop have a greater impact on its deformation compared to the viscoelastic properties of the surrounding ambient fluid. Specifically, we determine that the contribution of the $O(CDe)$ deformation term, denoted as $D^{(CDe)}$, becomes zero when the conductivity ratio $Q$ is equal to the permittivity ratio $R$. However, this contribution increases as the conductivity ratio decreases with a fixed permittivity ratio. We also find that viscoelasticity reduces prolate deformation; however, it enhances the magnitude of oblate deformation.

The transient dynamics of the drops exhibit distinct oscillatory patterns—a characteristic behavior of viscoelastic fluids. This behavior is attributed to the nonlinear nature of the Oldroyd-B constitutive model, which leads to a finite first normal stress difference during flow. In steady-state conditions, viscoelasticity significantly affects deformation. We represent our findings through a plot of normalized deformation against $De$.

In the $V_dN_m$ configuration, we observe a consistent trend where the overall deformation of the drop decreases as the Deborah number increases. This phenomenon is largely attributed to the viscoelastic normal stress at the poles, which reduces the axial length of the drop. In contrast, the viscoelastic tangential stress at the equator increases the length perpendicular to the axis. This interplay between normal and tangential stresses results in a net reduction in deformation with increasing Deborah numbers. When analyzing the $N_dV_m$ configuration, we observe a similar trend of decreasing deformation with increasing $De$, along with some non-monotonic behavior. In this case, the viscoelastic tangential stress at the equator increases the length perpendicular to the axis, while the viscoelastic normal stresses at the poles decrease the length. Interestingly, this viscoelastic normal stress can sometimes vary in a non-monotonic manner, resulting in an overall non-monotonic reduction in deformation with respect to $De$.

In conclusion, our investigation paves the way for further studies, particularly on drop breakup. We believe these results yield valuable insights into the stability and dynamics of viscoelastic fluid systems, enhancing our understanding of the complex behavior of drops in electric fields.

\appendix
\section{Asymptotic solution of drop subjected to electric field}\label{appA}

This appendix presents the asymptotic solution for an Oldroyd-B drop suspended in an Oldroyd-B medium, subjected to a steady, uniform electric field in the limit of small $C$ and small $De$. By substituting expanded physical quantities into the governing equations and boundary conditions and collecting terms at orders $O(1)$, $O(C)$, and $O(De)$, we obtain the governing equations and boundary conditions at the respective orders. This appendix details the solution of these equations and the calculation of the resulting drop deformation. The $O(1)$, $O(C)$, and $O(De)$ solutions for the electric and flow fields lead to drop deformations at orders $O(C)$, $O(C^2)$, and $O(CDe)$, respectively.

\subsection{Electric field at \texorpdfstring{$O(1)$}{O(1)}}

\noindent
Governing equations for the electric field at $O(1)$ are
\begin{equation}
    \grad^2 {\phi^{(0)}} = 0 \quad ; \quad {\bm{E}^{(0)}} = -\grad{\phi^{(0)}} \; .
\end{equation}
Governing equations are subjected to the following boundary conditions
\begin{align}
    \label{eqn_E0_bc_1}
    &\frac{\partial \phi^{(0)}_{i}}{\partial r} = 0 \quad \text{at} \quad r = 0 \quad \text{and} \quad
    \phi^{(0)}_{e} \rightarrow r\cos\theta \quad \text{as} \quad r \rightarrow \infty \; ;\\
    \label{eqn_E0_bc_2}
    &E^{(0)}_{i_\theta} = E^{(0)}_{e_\theta} \quad \text{and} \quad
    R E^{(0)}_{i_r} = E^{(0)}_{e_r} \quad \text{at} \quad r = 1 \;
\end{align}
where $i$ and $e$ are subscripts for drop and ambient phases, respectively. Subscripts $r$ and $\theta$ denote the components along radial and polar directions, respectively. The general solution of the Laplace equation in a spherical axisymmetric coordinate system is given by
\begin{equation}
    \phi^{(0)} = \sum_n \left( \bar{A}^{(0)E}_n r^n + \frac{\bar{B}^{(0)E}_n}{r^{n+1}} \right) P_n(\cos\theta) \;.
\end{equation}
Implementing the boundary conditions stated by \Cref{eqn_E0_bc_1}, the only mode of solution that remains is $P_{1}(\cos\theta)$. Hence, electric potentials for drop and ambient phases are given by
\begin{equation}
    \phi^{(0)}_{i} = {\bar{A}^{(0)E}_{i_1}} r \cos\theta \quad ; \quad
    \phi^{(0)}_{e} = \left(r + \frac{{\bar{B}^{(0)E}_{e_1}}}{r^{2}} \right) \cos\theta \;. 
\end{equation}
Implementing the boundary conditions \Cref{eqn_E0_bc_2} at the interface, constants are found as follows
\begin{equation}
    {\bar{A}^{(0)E}_{i_1}} = \frac{-3}{R + 2} \; ; \quad  
    {\bar{B}^{(0)E}_{e_1}} = \frac{R - 1}{R + 2}.
\end{equation}
Stress due to the electric field at $O(1)$ is calculated by Maxwell's stress tensor.
\begin{align}
\begin{rcases}
    \tau^{(0)E}_{i_{rr}} &= \frac{9 Q \cos2\theta}{2 (R+2)^2} \; ; \quad 
    \tau^{(0)E}_{i_{r\theta}} = -\frac{9Q\sin \theta \cos\theta}{(R+2)^2}\;;\\
    \begin{split}
    \tau^{(0)E}_{e_{rr}} & =  \frac{3 (R-1)(R - 1 + 2r^3) }{4 r^6 (R+2)^2} + \cos2\theta \left(\frac{5 (R-1)^2 + 2r^3\left(R^2+R-2\right)}{4 r^6 (R+2)^2} + \frac{1}{2}\right)
    \end{split}\;;\\
    \tau^{(0)E}_{e_{r\theta}} &= \sin2\theta \left(\frac{ (R-1)^2}{r^6 (R+2)^2} - \frac{ (R-1)}{2 r^3 (R+2)} - \frac{1}{2}\right)\;. 
\end{rcases}
\end{align}
Net electric stress at the spherical interface is given by
\begin{align}
    \tau^{(0)E}_{I_{rr}}  = \frac{9 \left(\left(1 - 2Q + R^2\right) \cos 2\theta+R^2 - 1\right)}{4 (R+2)^2} \quad ; \quad
    \tau^{(0)E}_{I_{r\theta}}  = \frac{9  (Q-R) \sin \theta \cos \theta}{(R+2)^2}.
    \label{eqn:electric_stress_1}
\end{align}

\subsection{Flow field at \texorpdfstring{$O(1)$}{O(1)}}

\noindent
At $O(1)$, the stress tensor reduces to the Newtonian stress tensor, leading to Newtonian flow equations
\begin{align}
    &\bm{\nabla}.{\bm{u}^{(0)}_{i}} = 0 \quad ; \quad -\bm{\nabla}{p^{(0)}_{i}} + \mu_r\bm{\nabla}^2{\bm{u}^{(0)}_{i}} = 0 \\
    &\bm{\nabla}.{\bm{u}^{(0)}_{e}} = 0 \quad ; \quad -\bm{\nabla}{p^{(0)}_{e}} + \bm{\nabla}^2{\bm{u}^{(0)}_{e}} = 0.
\end{align}
Since the flow is incompressible and axisymmetric, velocity components can be represented in terms of the Stokes stream function in spherical coordinates as follows
\begin{equation}
    {u^{(0)}_{r}} = \frac{1}{r^2 \sin\theta}\frac{\partial {\Psi^{(0)}}}{\partial \theta} \quad ; \quad {u^{(0)}_{\theta}} = -\frac{1}{r \sin\theta}\frac{\partial {\Psi^{(0)}}}{\partial r}.
\end{equation}
The governing equation for flow in terms of stream function is given by
\begin{equation}
    \bigg[\frac{\partial^2 }{\partial r^2} + \frac{\sin\theta}{r^2}\frac{\partial }{\partial \theta}\bigg(\frac{1}{\sin\theta}\frac{\partial }{\partial \theta}\bigg)\bigg]^2{\Psi^{(0)}} = 0.
\end{equation}
Boundary conditions on the flow field at $O(1)$ are given by
\begin{align}
    &{u^{(0)}_{e_r}}, \quad {u^{(0)}_{e_\theta}} = 0 \quad \text{as} \quad r \rightarrow \infty \quad \text{and} \quad
    \frac{\partial {u^{(0)}_{i_r}}}{\partial r}, \quad \frac{\partial {u^{(0)}_{i_\theta}}}{\partial r} = 0 \quad \text{at} \quad r = 0 \; ;\\
    &{u^{(0)}_{i_r}} = {u^{(0)}_{e_r}} = 0 \quad \text{and} \quad
    {u^{(0)}_{i_\theta}} = {u^{(0)}_{e_\theta}} \quad \text{at} \quad r = 1 \; ;\\
    &{\tau^{(0)H}_{i_{r\theta}}} - {{\tau}^{(0)H}_{e_{r\theta}}} = {\tau^{(0)E}_{I_{r\theta}}} \quad \text{at} \quad r = 1 \; ;\\
    &(-{p^{(0)}_{i}} + {\tau^{(0)H}_{i_{rr}}} + {\tau^{(0)E}_{i_{rr}}}) - (-{p^{(0)}_{e}} + {\tau^{(0)H}_{e_{rr}}} + {\tau^{(0)E}_{e_{rr}}}) = 2 f^{(C)} + \cot\theta f^{\prime{(C)}} + f^{\prime\prime{(C)}} \; .
\end{align}
The non-zero tangential electric stress acting on the interface can only be balanced by viscous stresses, leading to the development of a flow field inside and outside the drop. Consequently, the stream function should be chosen such that the resulting hydrodynamic tangential stress exhibits the same functional form as the electric tangential stress. From \Cref{eqn:electric_stress_1}, shear stress due to the electric field at the interface is of the form $f(r)\sin\theta\cos\theta$. Thus, the stream function is chosen as
\begin{equation}
    \Psi^{(0)} = g(r)\sin^2\theta\cos\theta \; .
\end{equation}
Substituting this form of the stream function in the governing equation, a fourth-order ordinary differential equation is obtained for the function $g(r)$, which gives the functional form for $g(r)$ as
\begin{equation}
    g(r) = \frac{\bar{A}}{r^2} + \bar{B} + \bar{C} r^3 + \bar{D} r^5 \;.
\end{equation}
Considering the boundary conditions far away from the drop and at the drop center, stream functions for drop and ambient phases are given as
\begin{align}
\begin{rcases}
    {\Psi^{(0)}_{i}} &= \left(\bar{C}^{(0)}_i r^3 + \bar{D}^{(0)}_i r^5\right)\sin^2\theta\cos\theta \; ;\\
    {\Psi^{(0)}_{e}} &= \left(\frac{\bar{A}^{(0)}_e}{r^2} + \bar{B}^{(0)}_e\right)\sin^2\theta\cos\theta \;.
\end{rcases}
\end{align}
Constants in the above expressions are obtained by implementing velocity continuity and tangential stress continuity at the interface.
\begin{align}
    \bar{A}^{(0)}_e = -\bar{B}^{(0)}_e = \bar{C}^{(0)}_i = -\bar{D}^{(0)}_i = U_0 =\frac{9 (Q-R)}{10 (\mu_r+1) (R+2)^2} \;.
\end{align}
The velocity field obtained by substituting the above constants in the general solution does not satisfy normal stress conditions at the spherical interface, and the interface of the drop undergoes deformation. $O(C)$ correction to the interface shape ($f^{(C)}(\theta)$) is obtained by implementing the normal stress boundary condition. Correction $f^{(C)}$ is given by
\begin{equation}
    f^{(C)} = \bar{B}_S^{(C)} P_2(\cos\theta)
\end{equation}
where,
\begin{equation}
    \bar{B}^{(C)}_S = \dfrac{3}{4(2 + R)^2}\bigg[1 - 2Q + R^2 + \dfrac{3}{5}(R - Q) \bigg(\dfrac{3\mu_r + 2}{\mu_r + 1}\bigg)  \bigg] \;.
\end{equation}

\subsection{Electric field at \texorpdfstring{$O(C)$}{O(C)}}

\noindent
Governing equations for electric field at $O(C)$ are similar to those for $O(1)$ as given by
\begin{align}
    &\bm{\nabla}^2 {\phi^{(C)}} = 0 \quad ; \quad {\bm{E}^{(C)}} = -\bm{\nabla}{\phi^{(C)}} \; .
\end{align}
However, they are subjected to the following boundary conditions
\begin{equation}
    \frac{\partial {\phi^{(C)}_{i}}}{\partial r} = 0 \quad \text{at} \quad r = 0 \; ;
\end{equation}
\begin{equation}
    {\bm{E}^{(C)}_{e}} \rightarrow \textbf{0} \quad \text{at} \quad r \rightarrow \infty \; ;
\end{equation}
\begin{equation}
    {E^{(C)}_{i_\theta}} + f^{(C)}\frac{\partial {E^{(0)}_{i_\theta}}}{\partial r} + {E^{(0)}_{i_r}}{f}^{\prime (C)} = {E^{(C)}_{e_\theta}} + f^{(C)}\frac{\partial {E^{(0)}_{e_\theta}}}{\partial r} + {E^{(0)}_{e_r}}{f}^{\prime (C)} \quad \text{at} \quad r = 1 \; ;
\end{equation}
\begin{equation}
    R\left({E^{(C)}_{i_r}} + f^{(C)}\frac{\partial {E^{(0)}_{i_r}}}{\partial r} - {f}^{\prime(C)}{E^{(0)}_{i_\theta}}\right) = {E^{(C)}_{e_r}} + f^{(C)}\frac{\partial {E^{(0)}_{e_r}}}{\partial r} - {f}^{\prime (C)}{E^{(0)}_{e_\theta}} \quad \text{at} \quad r = 1 \; .
\end{equation}
The general solution of the Laplace equation is given by
\begin{equation}
    \phi^{(C)} = \sum_n \left( {\bar{A}^{(C)E}_n} r^n + \frac{{\bar{B}^{(C)E}_n}}{r^{n+1}} \right) P_n(\cos\theta) \; .
\end{equation}
Only the first and third modes of solution remain to satisfy the boundary conditions. Implementation of boundary conditions leads to the following expressions for the constants
\begin{align}
\begin{rcases}
    \bar{A}^{(C)E}_{i_1} = -\frac{18 \bar{B}^{(C)}_S (R - 1)}{5(2 + R)^2} \quad ; \quad &\bar{A}^{(C)E}_{i_3} = 0 \; ;\\
    \bar{B}^{(C)E}_{e_1} = \frac{6 \bar{B}^{(C)}_S (R - 1)^2}{5(2 + R)^2} \quad ; \quad & \bar{B}^{(C)E}_{e_3} = \frac{9 \bar{B}^{(C)}_S (R - 1)^2}{5(2 + R)^2} \; .
\end{rcases}
\end{align}
Tangential stress due to the electric field at the interface at $O(C)$ is given by the following expression
\begin{equation}
    \tau^{(C)E}_{I_{nt}}= \bar{A}^{(C)E}_{I_{nt}}\sin2\theta + \bar{B}^{(C)E}_{I_{nt}}\sin4\theta \; ;
\end{equation}
where,
\begin{equation}
    \quad \quad \quad \quad \bar{A}^{(C)E}_{I_{nt}} = \frac{54 \bar{B}_S^{(C)} (R-1) (Q-R)}{5 (R+2)^3} \quad ; \quad
    \bar{B}^{(C)E}_{I_{nt}} = \frac{27 \bar{B}_S^{(C)} (Q-R)}{4(R+2)^2} \; .
\end{equation}
Normal stress at the interface due to the electric field is given by
\begin{equation}
    \tau^{(C)E}_{I_{nn}}= \bar{A}^{(C)E}_{I_{nn}} + \bar{B}^{(C)E}_{I_{nn}} P_2(\cos\theta) + \bar{C}^{(C)E}_{I_{nn}} P_4(\cos\theta)
\end{equation}
where,
\begin{align}
\begin{rcases}
    &\bar{A}^{(C)E}_{I_{nn}} = \frac{54 \bar{B}^{(C)}_S  (R+1) (Q-R)}{5 (R+2)^3} \; ;\\
    &\bar{B}^{(C)E}_{I_{nn}} = \frac{54 \bar{B}^{(C)}_S  (3 R-8) \left(1 - 2Q + R^2\right)}{35 (R+2)^3} \; ;\\
    &\bar{C}^{(C)E}_{I_{nn}} = \frac{216 \bar{B}^{(C)}_S  \left(1 - 2Q + R^2\right)}{35 (R+2)^2} \; .
\end{rcases}
\end{align}
Here, subscripts $nt$ and $nn$ represent the components of tensor in the coordinate system having unit vectors normal and tangential to the deformed interface, $\hat{\bm{n}}$ and $\hat{\bm{t}}$ as coordinate axes.

\subsection{Flow field at \texorpdfstring{$O(C)$}{O(C)}}

\noindent
Flow equations at $O(C)$ are
\begin{align}
    &\bm{\nabla}.{\bm{u}^{(C)}_i} = 0 \quad ; \quad -\bm{\nabla}{p^{(C)}_i} + \mu_r\bm{\nabla}^2{\bm{u}^{(C)}_i} = 0 \; ;\\
    &\bm{\nabla}.{\bm{u}^{(C)}_{e}} = 0 \quad ; \quad -\bm{\nabla}{p^{(C)}_{e}} + \bm{\nabla}^2{\bm{u}^{(C)}_{e}} = 0 \; ;
\end{align}
and they are subjected to the following boundary conditions
\begin{equation}
    {\bm{u}^{(C)}_{e}} = 0 \quad \text{as} \quad r \rightarrow \infty \quad \text{and} \quad 
    \frac{\partial {\bm{u}^{(C)}_{i}}}{\partial r} = 0 \quad \text{at} \quad r = 0 \; ;
\end{equation}
\begin{equation}
    {u^{(C)}_{i_r}} + f^{(C)}\frac{\partial {u^{(0)}_{i_r}}}{\partial r} - {f}^{\prime(C)}{u^{(0)}_{i_\theta}} = {u^{(C)}_{e_r}} + f^{(C)}\frac{\partial {u^{(0)}_{e_r}}}{\partial r} - {f}^{\prime(C)}{u^{(0)}_{e_\theta}} = 0 \quad \text{at} \quad r = 1  \; ;
\end{equation}
\begin{equation}
    {u^{(C)}_{i_\theta}} + f^{(C)}\frac{\partial {u^{(0)}_{i_\theta}}}{\partial r} + {f}^{\prime(C)}{u^{(0)}_{i_r}} =  {u^{(C)}_{e_\theta}} + f^{(C)}\frac{\partial {u^{(0)}_{e_\theta}}}{\partial r} + {f}^{\prime(C)}{u^{(0)}_{e_r}} \quad \text{at} \quad r = 1 \; ;
\end{equation}
\begin{multline}
    {\tau^{(C)H}_{i_{r\theta}}} + f^{(C)}\frac{\partial {\tau^{(0)H}_{i_{r\theta}}}}{\partial r} + {f}^{\prime(C)}({\tau^{(0)H}_{i_{rr}}} - {\tau^{(0)H}_{i{\theta\theta}}}) + {\tau^{(0)E}_{i_{nt}}}  \\= {\tau^{(C)H}_{e_{r\theta}}} + f^{(C)}\frac{\partial {\tau^{(0)H}_{e_{r\theta}}}}{\partial r} + {f}^{\prime(C)}({\tau^{(0)H}_{e_{rr}}} - {\tau^{(0)H}_{e_{\theta\theta}}}) + {\tau^{(0)E}_{e_{nt}}} \quad \text{at} \quad r = 1 \; ;
\end{multline}
\begin{multline}
    \bigg[- \bigg({p^{(C)}_{i}} + f^{(C)}\frac{\partial {p^{(0)}_{i}}}{\partial r}\bigg) + {\tau^{(C)H}_{i_{rr}}} + f^{(C)}\frac{\partial {\tau^{(0)H}_{i_{rr}}}}{\partial r} - 2{f}^{\prime(C)}{\tau^{(0)H}_{i_{r\theta}}} + {\tau^{(C)E}_{i_{nn}}}\bigg] \\- \bigg[- \bigg({p^{(C)}_{e}} + f^{(C)}\frac{\partial {p^{(0)}_{e}}}{\partial r}\bigg) + {\tau^{(C)H}_{e_{rr}}} + f^{(C)}\frac{\partial {\tau^{(0)H}_{e_{rr}}}}{\partial r} - 2{f}^{\prime(C)}{\tau^{(0)H}_{e_{r\theta}}} + {\tau^{(C)E}_{e_{nn}}}\bigg] \\= -\{2(f^{(C)})^2 - 2f^{(C^2)} + 2\cot\theta f^{(C)}f^{\prime(C)} - 2\cot\theta f^{\prime(C^2)} + 2f^{(C)}f^{\prime\prime(C)} \\- f^{\prime\prime(C)}\} \; .
\end{multline}
Under the assumption of axisymmetry, velocity can be expressed in terms of stream function, and the equation for stream function is given by,
\begin{align}
    \bigg[\frac{\partial^2 }{\partial r^2} + \frac{\sin\theta}{r^2}\frac{\partial }{\partial \theta}\bigg(\frac{1}{\sin\theta}\frac{\partial }{\partial \theta}\bigg)\bigg]^2\Psi^{(C)} = 0 \; .
\end{align}
From the tangential electric stress, the stream function should be chosen as $\Psi = (g(r) + h(r) \cos2\theta)\sin^2\theta\cos\theta$.
Substituting the stream function form in the governing equation and solving the resulting ordinary differential equations, general solutions of stream function equations are obtained. Applying boundary conditions far away and at the center of the drop, stream functions for drop and ambient phases are given by,
\begin{equation}
    {\Psi^{(C)}_{i}} = \bigg[\bigg(\frac{\bar{D}^{(C)}_{i(h)} r^7}{7} + \bar{C}^{(C)}_{i(g)} r^3 + \bar{D}^{(C)}_{i(g)} r^5\bigg) + \cos2\theta \bigg(\bar{C}^{(C)}_{i(h)} r^5 + \bar{D}^{(C)}_{i(h)} r^7\bigg)\bigg]\sin^2\theta\cos\theta \; ;
\end{equation}
\begin{equation}
    {\Psi^{(C)}_{e}} = \bigg[\bigg(\frac{\bar{A}^{(C)}_{e(h)}}{7r^4} + \frac{\bar{A}^{(C)}_{e(g)}}{r^2} + \bar{B}^{(C)}_{e(g)}\bigg) + \cos2\theta\bigg(\frac{\bar{A}^{(C)}_{e(h)}}{r^2} + \frac{\bar{B}^{(C)}_{e(h)}}{r^2} \bigg)\bigg]\sin^2\theta\cos\theta \; .
\end{equation}
After implementing boundary conditions at the interface, constants are found as follows
\begin{align}
    \bar{A}^{(C)}_{e(g)} = \frac{1}{420} \bar{B}^{(C)}_S U_0 \left(\frac{144}{\mu_r+1} - \frac{3024}{R+2} + 941\right)  \quad ; \quad &
    \bar{A}^{(C)}_{e(h)} = \frac{41 \bar{B}^{(C)}_S U_0}{12} \; ;\\
    \bar{B}^{(C)}_{e(g)} = \frac{6}{35} \bar{B}^{(C)}_S U_0 \left(-\frac{2}{\mu_r+1} + \frac{42}{R+2} - 13\right) \quad ; \quad &
    \bar{B}^{(C)}_{e(h)} = -\frac{23 \bar{B}^{(C)}_S U_0}{12} \; ;\\
    \bar{C}^{(C)}_{i(g)} = \frac{3 \bar{B}^{(C)}_S U_0 (21 \mu_r (R-2)+25 R-34)}{35 (\mu_r+1) (R+2)} \quad ; \quad &
    \bar{C}^{(C)}_{i(h)} = \frac{8 \bar{B}^{(C)}_S U_0}{3} \; ;\\
    \bar{D}^{(C)}_{i(g)} = \frac{\bar{B}^{(C)}_S U_0}{105} \left(-\frac{36}{\mu_r+1} + \frac{756}{R+2} - 119\right) \quad ; \quad &
    \bar{D}^{(C)}_{i(h)} = -\frac{7 \bar{B}^{(C)}_S U_0}{6}  \; .
\end{align}
Correction to the interface at O(${C}^2$) is found by normal stress conditions at the interface
\begin{equation}
    f^{(C^2)} = \bar{A}^{(C^2)}_S + \bar{B}^{(C^2)}_S P_2(\cos\theta) + \bar{C}^{(C^2)}_S P_4(\cos\theta)
\end{equation}
where,
\begin{multline}
    \bar{B}^{(C^2)}_S = \frac{2}{315} \bar{B}^{(C)}_SU_0 \bigg(\frac{150 \bar{B}^{(C)}_S}{U_0} + \frac{90 (\mu_r + 1) (3 R-8) \left(1 - 2 Q +R^2\right)}{(R + 2)(Q - R)} \\+ \frac{756 (3 \mu_r+2)}{R+2} - 897 \mu_r + \frac{36}{\mu_r+1}-456\bigg) \; ;
\end{multline}
\begin{equation}
    \bar{C}^{(C^2)}_S = \frac{8}{945} \bar{B}^{(C)}_SU_0 \left(\frac{45 \bar{B}^{(C)}_S}{U_0}-\frac{60 (\mu_r+1) \left(2 Q-R^2-1\right)}{Q-R}-113 \mu_r-88\right) 
\end{equation}
and from the volume conservation condition
\begin{equation}
    \bar{A}^{(C De)}_{S} = -\frac{1}{5}\bigg(\frac{4}{3}\bar{B}^{(C)}_S\bigg)^2 \; .
\end{equation}

\subsection{Flow field at \texorpdfstring{$O(De)$}{O(De)}}

\noindent
At $O(De)$, electric potential is governed by the Laplace equation. From boundary conditions, the electric field should go to zero far away from the drop, and the electric potential should be bounded at the drop center. Since both equations and boundary conditions are homogeneous, the electric field at $O(De)$ is zero. Hence, there is no electric stress at the interface at $O(De)$. However, flow equations at $O(De)$ have extra terms appearing in the stress tensor, which gives rise to an inhomogeneous part in the stream function equation. Governing equations for flow at $O(De)$, are given by,
\begin{align}
\begin{rcases}
    &\bm{\nabla}.{\bm{u}^{(De)}_{i}} = 0 \quad ; \quad -\bm{\nabla}{p^{(De)}_{i}} + \bm{\nabla}.{\bm{\tau}^{(De)H}_{i}} = 0 \; ;\\
    &\bm{\nabla}.{\bm{u}^{(De)}_{e}} = 0 \quad ; \quad -\bm{\nabla}{p^{(De)}_{e}} + \bm{\nabla}.{\bm{\tau}^{(De)H}_{e}} = 0 
\end{rcases}
\end{align}
where,
\begin{align}
    \begin{rcases}
    {\bm{\tau}^{(De)}_{i}} = \mu_r(\nabla {\bm{u}^{(De)}_{i}} + (\nabla {\bm{u}^{(De)}_{i}})^T) - \Lambda_i \Big[&{\bm{u}^{(0)}_{i}}.\bm{\nabla}{\bm{\tau}^{(0)Hp}_{i}}
    - {\bm{\tau}^{(0)Hp}_{i}}\bm{\nabla}{\bm{u}^{(0)}_{i}}\\ &- (\bm{\nabla}{\bm{u}^{(0)}_{i}})^T {\bm{\tau}^{(0)Hp}_{i}}\Big] \; ; \\
    {\bm{\tau}^{(De)}_{e}} = (\nabla {\bm{u}^{(De)}_{e}} + (\nabla {\bm{u}^{(De)}_{e}})^T) - \Lambda_e\Big[&{\bm{u}^{(0)}_{e}}.\bm{\nabla}{\bm{\tau}^{(0)Hp}_{e}}
    - {\bm{\tau}^{(0)Hp}_{e}}\bm{\nabla}{\bm{u}^{(0)}_{e}}\\ &- (\bm{\nabla}{\bm{u}^{(0)}_{e}})^T {\bm{\tau}^{(0)Hp}_{e}}\Big] \; .
    \end{rcases}
\end{align}
Governing equations at $O(De)$ are subjected to the following boundary conditions
\begin{align}
    &{\bm{u}^{(De)}_{e}} = 0 \quad \text{as} \quad r \rightarrow \infty \quad \text{and} \quad
    \frac{\partial {\bm{u}^{(De)}_{i}}}{\partial r} = 0 \quad \text{at} \quad r = 0 \; ;\\
    &{u^{(De)}_{i_r}} = {u^{(De)}_{e_r}} = 0 \quad \text{at} \quad r = 1 \quad \text{and} \quad
    {u^{(De)}_{i_\theta}} = {u^{(De)}_{e_\theta}} \quad \text{at} \quad r = 1 \; ;\\
    &{\tau^{(De)H}_{i_{r\theta}}} = {{\tau}^{(De)H}_{e_{r\theta}}} \quad \text{at} \quad r = 1 \; ;\\
    &(-{p^{(De)}_{i}} + {\tau^{(De)H}_{i_{rr}}}) - (-{p^{(De)}_{e}} + {\tau^{(De)H}_{e_{rr}}}) = 2 f^{(CDe)} + \cot\theta f^{\prime{(CDe)}} + f^{\prime\prime{(CDe)}} \; \text{at} \; r = 1 \; .
\end{align}
Equations for the stream functions are given by
\begin{multline}
    \bigg[\frac{\partial^2 }{\partial r^2} + \frac{\sin\theta}{r^2}\frac{\partial }{\partial \theta}\bigg(\frac{1}{\sin\theta}\frac{\partial }{\partial \theta}\bigg)\bigg]^2{\Psi^{(De)}_{i}} \\= - \frac{\Lambda_i}{\mu_r} r\sin\theta[ \bm{\nabla}\times(\bm{\nabla}.({\bm{u}^{(0)}_{i}}.\bm{\nabla}{\bm{\tau}^{(0)Hp}_{i}} - {\bm{\tau}^{(0)Hp}_{i}} \bm{\nabla}{\bm{u}^{(0)}_{i}} - (\bm{\nabla}{\bm{u}^{(0)}_{i}})^T {\bm{\tau}^{(0)Hp}_{i}} ))] \; ;
\end{multline}
\begin{multline}
    \bigg[\frac{\partial^2 }{\partial r^2} + \frac{\sin\theta}{r^2}\frac{\partial }{\partial \theta}\bigg(\frac{1}{\sin\theta}\frac{\partial }{\partial \theta}\bigg)\bigg]^2{\Psi^{(De)}_{e}} \\= -\Lambda_e r \sin\theta[\bm{\nabla}\times(\bm{\nabla}.({\bm{u}^{(0)}_{e}}.\bm{\nabla}{\bm{\tau}^{(0)Hp}_{e}} - {\bm{\tau}^{(0)Hp}_{e}}\bm{\nabla}{\bm{u}^{(0)}_{e}} - (\bm{\nabla}{\bm{u}^{(0)}_{e}})^T {\bm{\tau}^{(0)Hp}_{e}}))] \; .
\end{multline}
Constructing the right-hand side of the stream function equations from the flow field at $O(1)$, the following are the equations to be solved
\begin{equation}
    \bigg[\frac{\partial^2}{\partial r^2} + \frac{\sin\theta}{r^2}\frac{\partial }{\partial \theta}\bigg(\frac{1}{\sin\theta}\frac{\partial}{\partial}\bigg)\bigg]^2{\Psi^{(De)}_{i}} = -336 \Lambda_i(1 - \beta_i) U_0^2 r^3 \sin^2\theta\cos\theta \; ;
\end{equation}
\begin{multline}
    \bigg[\frac{\partial^2}{\partial r^2} + \frac{\sin\theta}{r^2}\frac{\partial }{\partial \theta}\bigg(\frac{1}{\sin\theta}\frac{\partial}{\partial}\bigg)\bigg]^2{\Psi^{(De)}_{e}} \\= 24(1 - \beta_e)\Lambda_eU_0^2\left(\left(\frac{35}{r^{9}} - \frac{9}{r^7}\right) + \cos2\theta\left(\frac{45}{r^{9}} - \frac{15}{r^7}\right)\right) \sin^2\theta\cos\theta \; .
\end{multline}
From the right-hand side of the stream function equation, the appropriate form for the stream function is chosen as $\Psi = (g(r) + h(r) \cos2\theta)\sin^2\theta\cos\theta$. Substituting the form of stream function in the governing equations and simplifying, differential equations are obtained for $g_{i}, h_{i}, g_{e}, h_{e}$. Solving those ordinary differential equations, the general solution for stream function equations is given as follows
\begin{multline}
    {\Psi^{(De)}_{i}} = \bigg[\bigg(\frac{2 (\beta_{i}-1) \Lambda_i r^7 U_0^2}{3} + \bar{C}^{(De)}_{i(g)}r^3 + \bar{D}^{(De)}_{i(g)}r^5 + \frac{\bar{D}^{(De)}_{i(h)}r^7}{7}\bigg) \\+ \left(\bar{C}^{(De)}_{i(h)} r^5 + \bar{D}^{(De)}_{i(h)} r^7\right)\cos 2\theta\bigg]\sin^2\theta\cos\theta \; ;
\end{multline}
\begin{multline}
    {\Psi^{(De)}_{e}} = \bigg[\bigg( -\frac{\Lambda_e(\beta_{e}-1)  U_0^2}{r^5} + \frac{\Lambda_e(\beta_{e}-1) U_0^2}{2 r^3} + \frac{\bar{A}^{(De)}_{e(g)}}{r^2} + \bar{B}^{(De)}_{e(g)}\bigg) \\ + \bigg(-\frac{9\Lambda_e(\beta_{e} - 1)  U_0^2}{2r^3} - \frac{3\Lambda_e(\beta_{e} - 1) U_0^2}{r^5} + \frac{\bar{A}^{(De)}_{e(h)}}{r^4} + \frac{\bar{B}^{(De)}_{e(h)}}{r^2} \bigg)\cos 2\theta\bigg]\sin^2\theta\cos\theta \; .
\end{multline}
Implementing the boundary conditions at the interface, the remaining constants are obtained as follows
\begin{align}
    \bar{A}^{(De)}_{e(g)} &= -\frac{U_0^2}{420 (\mu_r+1)} [196\mu_r\Lambda_i(\beta_i-1) + 3\Lambda_e(\beta_e-1)(25\mu_r + 98)] \; ;\\
    \bar{A}^{(De)}_{e(h)} &= \frac{U_0^2}{4 (\mu_r+1)} [-20\mu_r \Lambda_i  (\beta_i-1)  + \Lambda_e(\beta_e-1)(27\mu_r + 8)] \; ;\\
    \bar{B}^{(De)}_{e(g)} &= \frac{2 U_0^2}{105 (\mu_r+1)}  [62\mu_r\Lambda_i(\beta_i-1) - 3\Lambda_e(\beta_e-1)(5 \mu_r-16)] \; ;\\
    \bar{B}^{(De)}_{e(h)} &= \frac{U_0^2}{4 (\mu_r+1)} [20\mu_r\Lambda_i(\beta_i-1)   + \Lambda_e(\beta_e-1)(3 \mu_r+22)] \; ;\\
    \bar{C}^{(De)}_{i(g)} &= -\frac{2 U_0^2}{105 (\mu_r+1)} [\Lambda_i(\beta_i-1)(27 \mu_r-35) + 63\Lambda_e(\beta_e-1)]\; ;\\
    \bar{C}^{(De)}_{i(h)} &= -\frac{U_0^2}{4 (\mu_r+1)} [20\mu_r\Lambda_i (\beta_i-1) + 19\Lambda_e(\beta_e-1)]\; ;\\
    \bar{D}^{(De)}_{i(g)} &= \frac{U_0^2}{420 (\mu_r+1)} [- 28\Lambda_i(\beta_i-1)(13 \mu_r+20) + 219\Lambda_e(\beta_e-1) ] \; ;\\
    \bar{D}^{(De)}_{i(h)} &= \frac{U_0^2}{4 (\mu_r+1)} [20\mu_r\Lambda_i(\beta_i-1) + 19\Lambda_e(\beta_e-1) ] \; .
\end{align}
Correction to the interface shape at $O(CDe)$ is calculated by normal stress conditions at the interface.
\begin{equation}
    f^{(CDe)} = \bar{B}^{(CDe)}_SP_2(\cos\theta) + \bar{C}^{(CDe)}_SP_4(\cos\theta)
\end{equation}
where,
\begin{equation}
    \bar{B}^{(CDe)}_S = \frac{U_0^2}{105 (\mu_r+1)} [2\mu_r\Lambda_i (\beta_{i}-1)  (123 \mu_r+92) + \Lambda_e(\beta_e-1)  (79 \mu_r+16)] \; ;
\end{equation}
\begin{equation}
    \bar{C}^{(CDe)}_S = \frac{U_0^2}{105 (\mu_r+1)} [4\mu_r\Lambda_i(\beta_{i}-1)  (52 \mu_r+47)-\Lambda_e(\beta_e-1)  (101 \mu_r+120)] \; .
\end{equation}

\bibliographystyle{elsarticle-harv} 
\bibliography{references}



\end{document}